\documentclass[amsmath,amssymb,aps,pra,showpacs,reprint]{revtex4-1}
\usepackage[utf8]{inputenc}
\usepackage[english]{babel}
\usepackage{graphicx}
\usepackage{dcolumn}
\usepackage{bm}
\usepackage[breaklinks]{hyperref}
\usepackage{bbold}
\usepackage{color}
\usepackage[normalem]{ulem}
\usepackage{feynmp}
\renewcommand{\vec}[1]{\mathbf{#1}}

\begin{document}

\preprint{APS/123-QED}

\title{Equilibration and Approximate Conservation Laws: \\
Dipole Oscillations and Perfect Drag of Ultracold Atoms in a Harmonic Trap}

\author{Robert Bamler}
\author{Achim Rosch}
\affiliation{Institute for Theoretical Physics, Universit\"at zu K\"oln, D-50937 K\"oln, Germany}

\date{\today}

\begin{abstract} 
The presence of (approximate) conservation laws can prohibit the fast relaxation of interacting many-particle quantum systems.
We investigate this physics by studying the center-of-mass oscillations of two species of fermionic ultracold atoms in a harmonic trap.
If their trap frequencies are equal, a dynamical symmetry (spectrum generating algebra), closely related to Kohn's theorem, prohibits the relaxation of center-of-mass oscillations.
A small detuning $\delta\omega$ of the trap frequencies for the two species breaks the dynamical symmetry and ultimately leads to a damping of dipole oscillations driven by inter-species interactions.
Using memory-matrix methods, we calculate the relaxation as a function of frequency difference, particle number, temperature and strength of inter-species interactions. When interactions dominate, there is almost perfect drag between the two species and the dynamical symmetry is approximately restored. The drag can either arise from Hartree potentials or from friction. In the latter case (hydrodynamic limit), the center-of-mass oscillations  decay with a tiny rate, $1/\tau \propto (\delta\omega)^2/\Gamma$, where $\Gamma$ is a single particle scattering rate.
\end{abstract}

\maketitle

How does an interacting many-body quantum system reach thermal equilibrium? While often a few scattering processes are sufficient to establish locally an approximate equilibrium state, in some cases the presence of conservation laws prohibits equilibration.
In one dimension (1D), for example, integrable quantum systems like the spin 1/2 Heisenberg model or the fermionic Hubbard model possess an infinite number of conservation laws. Due to their presence, the system cannot relax to a simple thermal state described by just a few parameters like temperature or chemical potential. Instead, only an equilibration to a generalized Gibbs ensemble (GGE) \cite{rigol_relaxation_2007} is expected where for each conservation law a new Lagrange parameter is needed to describe the long-time steady state.

Real experimental systems are, however, often only approximately described by integrable models.
As a consequence the corresponding conservation laws are only approximately valid. For classical systems with a finite number of degrees of freedom, the famous KAM theorem \cite{kolmogorov_conservation_1954,arnold_proof_1963,moser1962invariant} states that even in such a situation many properties of the integrable point can survive. The situation for interacting many-particle quantum systems is less clear. Generally it is, however, expected that due to integrability breaking terms the system can relax to a thermal state but the relaxation is slow and governed by the slow relaxation of the approximate conservation laws.
A similar question arises in transport studies:
Integrable systems like the 1D Heisenberg model are characterized by infinite (heat-) conductivities even at finite temperature \cite{zotos_transport_1997,zotos_issues_2005,ford_statistical_1965}.
In real materials, however, small integrability breaking terms can render the conductivity finite.
This has motivated early studies of the role of integrability breaking terms for transport properties \cite{jung_transport_2006,jung07}.

Ultracold atoms provide new opportunities to investigate the question of equilibration and the role of (approximate) symmetries. For example, in a famous experiment termed ``Quantum Newton's Cradle''\cite{kinoshita_quantum_2006} it was shown that the breathing mode of a 1D Bose liquid in a harmonic trap does not relax on experimentally relevant time scales. While in this case the harmonic traps nominally break integrability this apparently has little effect on the experiments.

In this paper, we study equilibration in the presence of an approximate symmetry in a model which is (i)  ideally suited for experimental studies and (ii) conceptually simple due to the presence of only a single
symmetry -- instead of infinitely many. We study the center-of-mass (COM) oscillations of atoms in a harmonic trapping potential. If all atoms have the {\em same} mass and {\em same} trapping potential, then the COM oscillation never decays and its frequency is exactly given by the non-interacting result \cite{brey_optical_1989}. A closely related results is Kohn's theorem \cite{kohn_cyclotron_1961} stating that cyclotron resonances of electrons in a Galileian invariant system are not affected by interactions. Mathematically, this can be traced back to fact that the total momentum $\vec P$, the center-of-mass $\vec R$ and the interacting many-particle Hamiltionan in the presence of a trapping potential $\frac{1}{2} V_0 \vec r^2$ form a closed algebra (a so-called {\em spectrum generating algebra}) given by
\begin{align}
	[R^i,P^j]\!=\!i \hbar \delta_{ij}, \  [P^i,H]\!=\!-i \hbar V_0 N R^i, \  [R^i,H]\!=\! i \frac{\hbar}{N m} P^i
\end{align}
This algebra implies that the COM motion completely separates from all many-particle excitations in the trap even in the presence of a time-dependent trapping potential $\frac{1}{2} V_0(t) \vec (r-r_0(t))^2$.
Furthermore, the nonlocal (!) operator $Q={\vec P}^2/(2 m)+\frac{1}{2} V_0 {\vec R}^2$ is a conservation law, $[Q,H]=0$.
We will study how these symmetries break down when two species of atoms with slightly different masses or slightly different trapping potentials are considered.
Such a case has recently been studied by the Salomon group \cite{ferrierbarbut_mixture_2014} 
using mixtures of $^6$Li and $^7$Li. This work investigated, however, mainly  the role of superfluidity on the COM oscillations in this system.

The case of a fermionic mixture has been studied theoretically by Chiacchiera, Macr\`i and Trombettoni
\cite{chiacchiera_dipole_2010}.
While the methods used by the authors are similar to the one used in our study (projection on the dynamics of slow modes), their paper mainly focuses on counting the number of relevant modes and contains little information on the question discussed in this paper, especially on the behavior of the damping rate as function of the trap-frequency difference, population difference, inter-species scattering rate, and temperature.
Furthermore, it only considers the classical high-temperature limit where effects of Pauli blocking can be ignored.

An alternative option to perturb the dynamical symmetry of COM oscillations is to consider corrections to the confining harmonic potential, e.g., by adding a $r^4$ term. Such a situation has very recently be investigated for a quasi one-dimensional setup in the hydrodynamic limit by Iqbal, Levchenko and Khodas \cite{iqbal_decay_2015} using the Navier-Stokes equation. Similar to the case discussed in this paper, they obtain a long-lived mode where the decay rates are controlled by the strength of the anharmonic terms.

A spectrum generating algebra also characterizes approximately the breathing mode (monopole oscillations) of atoms in a harmonic trap in two dimensions \cite{pitaevskii_breathing_1997} and of a unitary gas in arbitrary dimensions \cite{castin}.
In this case, the shift of the resonance frequency due to deviations from the unitary limit has been calculated in one dimension in Ref.~ \cite{menotti_collective_2002,zhang_breakdown_2014} while the two-dimensional case was studied in \cite{olshanii_example_2010,hofmann_quantum_2012,taylor_apparent_2012} and investigated experimentally in \cite{vogt_scale_2012}. 
Within our study, we will be mainly interested to study the relaxation rate rather than the frequency shift.

The relative motion of two species of atoms is mainly controlled by their mutual interactions. This problem, often described by the term {\em spin drag}, has been investigated both in the context of electrons in solids \cite{spindragElectron,spindragElectron2} and also for ultracold atoms, see e.g. \cite{spindragUltra,bruun}.

In the following, we will first introduce the model and our analytical approach, identify three important physical regimes (ballistic, frictionless drag, and friction dominated drag), and, finally, quantitatively predict how
these regimes determine properties both for the real-time evolution and for the response as function of the frequency.

\section{Model and method}

In this article we study two species of ultracold fermions with creation operators $\Psi_1^\dagger(\vec r)$ and $\Psi_2^\dagger(\vec r)$ captured each in a perfectly harmonic trap in three dimensions. The system is described by
\begin{alignat}{1}\label{hamiltonian}
	H &= H_0 + H^{(11)}_{\text{int}} +H^{(22)}_{\text{int}}+ H^{(12)}_{\text{int}};  \\
	H_0 &= \sum_{i=1}^2 \int\!d^3r\; \Psi_i^\dagger(\vec r)
	       \!\left[-\frac{\hbar^2 \nabla^2}{2m_i} + \frac{m_i\omega_i^2}{2}(\vec r-\vec r_i^0)^2 \right]\!
	       \Psi_i(\vec r) \nonumber \\
	H^{(12)}_{\text{int}} &= \frac{4\pi\hbar^2a}{2m_{\text{red}}}
         \int\!d^3r\; \Psi_1^\dagger(\vec r) \Psi_2^\dagger(\vec r) \Psi_2(\vec r) \Psi_1(\vec r) \nonumber
\end{alignat}
Here, a shift of the position of the potential minimum $\vec{r}_i^0(t)$ can be used to excite dipolar oscillations.
In general, the two fermion species may have different masses $m_i$ and feel different trap potentials with respective trap frequencies $\omega_1=\bar \omega+\frac{\delta \omega}{2}$, $\omega_2=\bar \omega-\frac{\delta \omega}{2}$. $H^{(ii)}_{\text{int}}$ describes the intra-species interaction which we do not specify here as it does not influence our results in any qualitative way.
Furthermore, for spinless fermions  $H^{(ii)}_{\text{int}}$ can safely be neglected.
As we will show, all relaxation arises from the inter-species interaction which we parametrize by the s-wave scattering length  $a$ with $m_{\text{red}}=1/(m_1^{-1}+m_2^{-1})$ being the reduced mass (note that we use a pseudopotential to describe the scattering, see, e.g., Ref.~\cite{bloch_many-body_2008}).
For $\delta \omega=0$ the COM oscillations do not decay (see below).
We are therefore mainly interested in the limit $\delta \omega\ll \bar \omega$, where a slow decay of the oscillations can be expected. Experimentally, this can, for example, be realized by using two isotopes with slightly different mass, $m\pm\frac{\delta m}{2}$, but identical trapping potential.
In this case $\frac{\delta \omega}{\bar \omega}=-\frac{\delta m}{2 m}$.
Alternatively, one can use two hyperfine states of the same atom in combination with a spin-dependent potential \cite{bloch_many-body_2008}.
The latter setup has the advantage that one can directly tune the parameter $ \frac{\delta \omega}{\bar \omega}$.

Our theoretical approach is based on the idea that for  $\frac{\delta \omega}{\bar \omega} \ll 1$ the dynamics is governed by an approximate dynamical symmetry which 
prohibits a fast relaxation of the COM oscillations. Furthermore, in the limit of vanishing inter-species interactions, $a \to 0$, also the COM motion of each atomic species separately decouples.  Our central goal is to derive an effective, hydrodynamic description of the slowly relaxing modes. We will therefore focus on the dynamics in the operator space spanned by the  center-of-mass coordinates $\vec R_i$ and the total momentum $\vec P_i$ of each of the two species defined by
\begin{alignat}{1}\label{comcoordinates}
	\vec R_i &= \frac{1}{N_i} \int\!d^3r\; \Psi_i^\dagger(\vec r)\,\vec r\,\Psi_i(\vec r); \\
	\vec P_i &= \int\!d^3r\; \Psi_i^\dagger(\vec r)\,(-i\hbar\nabla)\,\Psi_i(\vec r)\nonumber
\end{alignat}
where $N_i$ is the number of particles of type $i=1,2$.

For weak excitations of the system, it is sufficient to study linear response within the Kubo formalism. 
The main goal is thereby to calculate the matrix of retarded susceptibilities 
\begin{eqnarray}\label{def-chi}
\chi_{mn}(\omega)= \frac{i}{\hbar}  \int_0^\infty\!dt\; e^{i \omega t}\langle [A_m(t),A_n(0)]\rangle_{\text{eq.}}
\end{eqnarray}
where $\langle\cdot\rangle_{\text{eq.}}$ denotes the expectation value in equilibrium for $\vec r_i^0(t)=0\;\forall t$ and $A_n=(R_1^x,R_2^x,P_1^x,P_2^x)$. As for a spherical potential the $x$, $y$ and $z$ components do not mix within linear response, we can focus on the $x$ coordinate only. $\chi_{mn}$ allows to calculate all experiments where the COM oscillations are excited by a shift $\vec r_i^{0}(t)$ of the potential and where the COM and/or the average momenta of the particles are observed.

To calculate $\chi_{mn}(\omega)$ we use the so-called memory matrix formalism \cite{forster_hydrodynamic_1995,mori_transport_1965,zwanzig_ensemble_1960}.
The memory matrix is a matrix of relaxation rates of slow variables, which we evaluate perturbatively in the strength of the inter-species interaction.
This formalism has the advantages that (i) it is easy to evaluate -- without the need to solve the type of integral equations needed for Boltzmann approaches or when vertex corrections are taken into account within the Kubo formalism, (ii) it nevertheless automatically includes the effect of vertex corrections, which are essential to describe momentum conservation, which is also governing the COM oscillations \cite{goetze}, (iii) it is accurate in cases where there is a separation of time scales and all slow modes are included in the memory matrix, (iv) it can be used to treat complicated situations like the expansion around a fully interacting integrable system \cite{jung_transport_2006,jung07} and has recently been used to calculate transport properties of exotic non-Fermi liquids \cite{hartnoll_transport_2014,lucas_scale-invariant_2014,lucas_memory_2015} (v) in the case considered here, where we effectively expand around the non-interacting limit, it is equivalent to a solution of the Boltzmann equation by projection onto the slow modes \cite{belitz_electronic_1984,chiacchiera_dipole_2010}.
In Ref.~\onlinecite{jung_lower_2007} we have argued that the formalism gives always a lower limit for conductivities. The situation investigated here is, however, more complicated compared to the case
considered in Ref.~\onlinecite{jung_lower_2007} as we are studying here effects at finite frequency in a system which is not translationally invariant. This leads to extra dephasing effects discussed in detail in Appendices \ref{appendix-sigma-singular} and \ref{appendix-dephasing}.

We refer to Appendix \ref{appendix-memorymatrix} for a brief review of the memory-matrix method.
It allows to express the matrix $\chi_{mn}(\omega)$ of retarded susceptibilities  (cf.~Eqs.(\ref{relationcchi},\ref{cofomegamemory})),
\begin{equation}\label{chiMemoryMatrix}
	\chi(\omega) = \left(1- \omega\left(\omega-\Omega + i\Sigma(\omega)\right)^{-1}\right) C_0
\end{equation}
in terms of an equal-time correlation matrix $C_0$, a constant matrix $\Omega$ and a frequency-dependent matrix-valued complex function $\Sigma(\omega)$. The latter two matrices have a similar role as the self-energy: they describe directly the
shift of frequencies and the damping of oscillations. They have the advantage that they can be evaluated
directly in perturbation theory, without the need to resum an infinite series of diagrams. More precisely,
the latter statement  holds in the case when all slow modes have been included in the set of observables $A_n$. We will use $A_n=(R_1^x,R_2^x,P_1^x,P_2^x)$ as the slow modes, which is sufficient to describe the regime where interactions dominate. As we discuss in detail in section \ref{appendix-sigma-singular} of the appendix, in the limit of vanishing interactions an infinite set of further slow modes exists, which have to be included to describe details
of the dephasing of oscillations for very weak interactions (ballistic regime) studied in detail in Appendix \ref{appendix-dephasing} but not captured for the above choice of $A_n$.

In the following our goal will be to calculate for weak interactions the frequencies and decay rates
of the center-of-mass oscillations.
In appendix \ref{appendix-evaluation-sigma}, we evaluate the matrices $\Omega$, $\Sigma(\omega)$, and $C_0$ in local density approximation for weak interactions.
To linear order in $a$, using Eqs.~\eqref{omega0} and \eqref{omega1}, we find $\Omega=\Omega^{(0)}+\Omega^{(1)}$ with
\begin{alignat}{1}\label{omega0Text}
	\Omega^{(0)} &= \begin{pmatrix}
		0 & 0 & i/M_1 & 0 \\
		0 & 0 & 0 & i/M_2 \\
		-iM_1 \omega_1^2	& 0 & 0 & 0 \\
		0 & -iM_2 \omega_2^2 & 0 & 0
	\end{pmatrix} 
\end{alignat}
and
\begin{alignat}{1}\label{deltaOmega}
	\Omega^{(1)} &= i  \gamma M_2 \omega_2 \begin{pmatrix}
		0 & 0 & 0 & 0 \\
		0 & 0 & 0 &0 \\
		1 & -1 & 0 & 0 \\
		-1 & 1 & 0 & 0
	\end{pmatrix}.
\end{alignat}
Here, $M_i=N_i m_i$ is the total mass of the fermions of species $i$ and $\gamma$ has the unit of a rate and is linear in the scattering length $a$ but depends in general on temperature and other parameters (see below).

Physically, Eq.~\eqref{omega0Text} describes independent oscillations of the two species in the absence of interactions.
The eigenfrequencies of $\Omega^{(0)}$ are given by the trap frequencies, $\pm \omega_1$ and $\pm \omega_2$.
Eq.~(\ref{deltaOmega}) describes that each species introduces a Hartree potential for the other species.
As we will discuss below, this contribution will shift the oscillation frequencies as long as the two species do not oscillate in parallel.
We obtain within a local density approximation using Eqs.~\eqref{omega1} and \eqref{r1r2scalarproduct} from the appendix,
\begin{alignat}{1}\label{gammaexact}
	\gamma &= \frac{a k_B T \omega_1^2 \omega_2}{3\pi\hbar^4} \frac{m_1^{5/2} m_2^{3/2}}{N_2 m_{\text{red}}}
		\int_0^\infty\!dr\, r^4 g_1(r)g_2(r)
\end{alignat}
with
\begin{alignat}{1}\label{gammaexactAux}
	g_i(r) &= \text{Li}_{\frac12}\!\left(-e^{(\mu_i - \frac12 m_i\omega_i^2 r^2)/(k_BT)}\right)
\end{alignat}
where $\text{Li}_{\frac12}$ is the polylogarithm of order $\frac12$ and $\mu_i$ is the chemical potential for particles of species $i$ in the limit $a\to0$, see the discussion in appendix \ref{appendix-eval-c0}.

Damping, described by $\Sigma(\omega)$, arises only to second order in the interaction strength.
The total momentum is conserved during scattering processes, $\partial_t {\bf P}_1=-\partial_t {\bf P}_2$, which leads to the simple matrix structure
\begin{alignat}{1}\label{mmatrix}
	 \Sigma(\omega\to  0) &\approx \Gamma \begin{pmatrix}
		0 & 0 & 0 & 0 \\
		0 & 0 & 0 & 0 \\
		0 & 0 & M_2/M_1 & -1 \\
		0 & 0 & -M_2/M_1 & 1
	\end{pmatrix}
\end{alignat}
with
\begin{alignat}{1}\label{Gammaexact}
	\Gamma= &\frac{\pi\hbar}{M_2 k_B T} \left(\frac{4\pi\hbar^2a}{2m_{\text{red}}}\right)^2
	\int\!d^3r
	\prod_{\substack{i=1,2;\\\alpha=1,2}} \int\!\frac{d^3k_{i\alpha}}{(2\pi)^3} \times \nonumber\\
	&\quad\times
	\delta(\Delta\epsilon) \delta^{(3)}(\Delta \vec k)
	q_x^2\, f_{11}f_{21}(1-f_{12})(1-f_{22})
\end{alignat}
to second order in the interaction strength using again the local density approximation, see Appendix \ref{appendixSigma}.
Here $f_{i\alpha}$ are Fermi functions evaluated at the energy $\epsilon_{i\alpha}=\hbar^2 k_{i \alpha}^2/(2 m_i)+\frac{1}{2} m_i \omega_i^2 {\bf r}^2$ and $\mathbf q=\mathbf k_{11}-\mathbf k_{12}$ is the change of momentum of the first species, while $\Delta \mathbf k$ and $\Delta \epsilon$ is the change of total momentum and energy, respectively.
As the oscillation frequency is assumed to be much smaller than all Fermi energies, we have used the limit $\omega \to 0$.
Furthermore, we ignore all frequency shifts to order $a^2$ (arising from the Kramers-Kronig partner of $\Gamma$). A more subtle issue is that our approach also neglects the coupling of the COM oscillations to other modes oscillating with frequency $\omega_i$ for $a \to 0$.
This is justified as, in the presence of interactions, these modes decay rapidly, but formally breaks down in the limit of vanishing interactions.
As discussed in more detail in the supplement, this approximation gives rise to small, but nominally divergent extra contribution to $\Sigma(\omega)$, which do, however, not affect our results.

Finally, the equal-time correlation matrix $C_0$ in Eq.~\eqref{chiMemoryMatrix} is evaluated in Appendix \ref{appendix-eval-c0}.
To linear order in $a$, we obtain $C_0=C_0^{(0)}+C_0^{(1)}$ where
\begin{alignat}{1}\label{c0}
	C_0^{(0)} &= \begin{pmatrix}
		1/(M_1 \omega_1^2) & 0 & 0 & 0 \\
		0 & 1/(M_2 \omega_2^2) & 0 & 0 \\
		0 & 0 & M_1 & 0 \\
		0 & 0 & 0 & M_2
	\end{pmatrix}
\end{alignat}
and
\begin{equation}
	C_0^{(1)} = \frac{\gamma}{M_1 \omega_1^2 \omega_2} \begin{pmatrix}
		\frac{M_2\omega_2^2}{M_1\omega_1^2} & -1 & 0 & 0 \\
		-1 & \frac{M_1\omega_1^2}{M_2\omega_2^2} & 0 & 0 \\
		0 & 0 & 0 & 0 \\
		0 & 0 & 0 & 0
	\end{pmatrix},
\end{equation}
where $\gamma\propto a$ is given in Eqs.~\eqref{gammaexact}--\eqref{gammaexactAux}.

\section{Analytic results\label{sectionAnalyticResults}}

Three different regimes have to be distinguished when discussing how the interactions affect the COM oscillations, depending on which of the three quantities $\delta \omega$, $|\gamma|$ and $\Gamma$ is largest.
First, in the {\em ballistic regime} ($\delta \omega \gg |\gamma|, \Gamma$) interaction effects can approximately be ignored and the oscillations of the two species are almost independent.
Second, in the {\em frictionless drag regime} ($|\gamma| \gg \delta \omega, \Gamma$) one species drags the other by the interaction-induced Hartree potential.
Finally, in the {\em friction dominated drag regime}  ($\Gamma \gg \delta \omega, |\gamma|$) the two clouds are coupled by friction and only a hydrodynamic COM oscillation with small effective damping survives.

\begin{figure}
\includegraphics[width=\columnwidth]{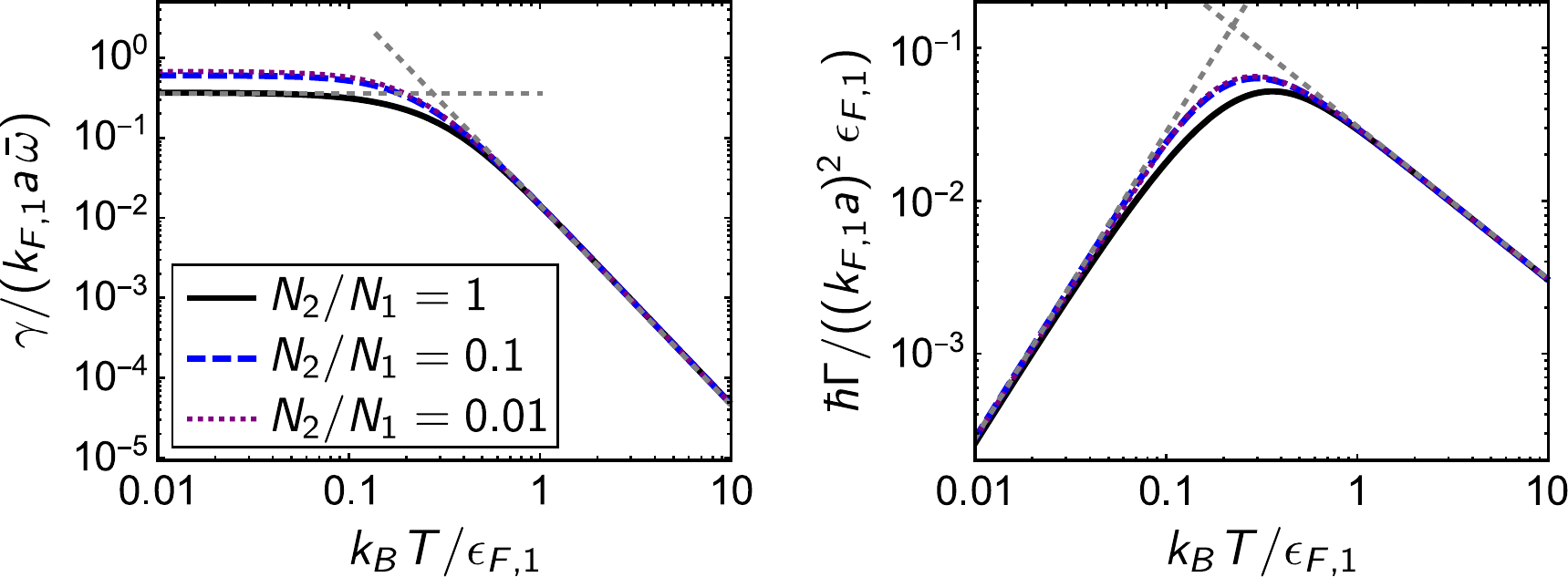}
\caption{\label{fig_gammaplot}
(Color online) Numerical results for the quantities $\gamma$ (Eqs.~\eqref{gammaexact}--\eqref{gammaexactAux}) and $\Gamma$ (Eq.~\eqref{Gammaexact}) for different ratios of $N_2/N_1$.
Dotted gray lines are analytic predictions for $T\ll\epsilon_{F,1}$ and $T\gg\epsilon_{F,1}$, see Eqs.~\eqref{lowTlimits}--\eqref{highTlimits}.
The analytic formula for $\gamma$ in the limit $T\to0$ given in Eq.~\eqref{lowTlimits} is only exact for $N_1=N_2$ and underestimates the value of $\gamma$ for $N_2/N_1\to0$ by a factor of $64/(35\pi)\approx0.58$.
All curves were calculated with $m_1=m_2$, $\frac{\delta\omega}{\bar\omega}=0.1$, and are independent of the total particle number in the chosen units.
}
\end{figure}

For a quantitative calculation we have evaluated the integrals in Eqs.~\eqref{gammaexact} and \eqref{Gammaexact} numerically, see section \ref{sectionNumericalResults} and Fig.~\ref{fig_gammaplot}.
In the limit of very low or very high temperature, also an analytic calculation is possible.
For low temperatures, $k_BT\ll\epsilon_{F,1}$ and $N_2 \leq N_1$ one finds
\begin{alignat}{1}
	\gamma &\approx \frac{128}{35\pi^2}\, k_{F,1} a\, \bar\omega \approx 0.37 \,k_{F,1} a\, \bar\omega \nonumber \\
	\Gamma &\approx \frac{8\pi}{9} \frac{(k_B T)^2}{\hbar\, \epsilon_{F,1}} \,(k_{F,1}a)^2
	\label{lowTlimits}
\end{alignat}
where $k_{F,i}$ is the Fermi momentum of species $i$ in the center of the trap with Fermi energy $\epsilon_{F,i}=k_{F,i}^2/(2 m_i)$ determined for $T \to 0$.
The analytic formulas have been computed in the limit $\delta \omega \to 0$ and for $m_1=m_2$.
While the prefactor of $\Gamma$ is valid for arbitrary ratios of $N_2$ and $N_1$ as long as $N_2\leq N_1$, the prefactor for $\gamma$ is only exact for $N_1=N_2$ but increases by less than a factor of 2 when $N_2/N_1$ is reduced, see Fig.~\ref{fig_gammaplot}.
Surprisingly, the estimates given in Eq.~\eqref{lowTlimits} are even valid when the temperature is larger than the Fermi energy of the second species.
If the temperature is larger than both Fermi energies, in contrast, the scattering rate $\Gamma$ drops with $1/T$ while $\gamma$ vanishes with $1/T^{5/2}$,
\begin{alignat}{1}
	\gamma &\approx \frac{k_{F,1}a\,\bar\omega}{24\sqrt{2\pi}} \left(\frac{\epsilon_{F,1}}{k_B T}\right)^{5/2} = \frac{k_T a\,\bar\omega}{24\sqrt{2\pi}} \left(\frac{\epsilon_{F,1}}{k_B T}\right)^3 \nonumber \\
	\Gamma &\approx \frac{(k_{F,1}a)^2}{9\pi} \frac{\epsilon_{F,1}^2}{\hbar\, k_B T} = \frac{(k_Ta)^2}{9\pi} \frac{\epsilon_{F,1}^3}{\hbar\, (k_B T)^2}
	\label{highTlimits}
\end{alignat}
where $k_T=\sqrt{2mk_BT}/\hbar$ is the thermal wave vector.
The prefactors for the high-temperature limit of both $\gamma$ and $\Gamma$ are valid for arbitrary ratios of $N_2$ and $N_1$ as long as $N_2\leq N_1$.

In both regimes, $\Gamma$ can be identified with the single-particle scattering rate of a particle of species $2$ in the center of the trap.
In the high-temperature regime, this can be seen by rewriting $\Gamma\sim \sigma v_{\rm th} n_1$ in terms of the scattering cross section $\sigma \sim a^2$, the typical velocity $ v_{\rm th} \sim \sqrt{k_B T /m}$ and the density of particles of species $1$ in the center, $n_1 \sim N_1/(T/m \omega^2)^{3/2} \sim \epsilon_{F,1}^3 (m/T)^{3/2}$.

\subsection{Ballistic regime}

For very small interactions the two species oscillate approximately independently of each other.
More precisely, we require that the strength $\gamma$ of the effective interaction potential and the single-particle scattering rate $\Gamma$ are both smaller (in magnitude) than the {\em difference} of oscillation frequencies, $|\gamma| \ll \delta \omega$ and $\Gamma \ll \delta\omega$.
While this regime is usually not realized experimentally at low temperatures (without tuning interactions close to zero), we discuss it here for completeness.
Note that this regime is always reached in the limit of high temperatures as long as $\delta\omega \neq 0$.

The remaining weak interactions lead to a small shift of the respective oscillation frequencies relative to the trap frequencies and to a finite, but long, lifetime of the two oscillatory modes.
The complex eigenfrequencies are given by the eigenvalues of $\Omega-i\Sigma(\omega)$, where the matrix of retarded susceptibilities, Eq.~\eqref{chiMemoryMatrix}, has poles.
We find for the eigenfrequencies in the ballistic regime,
\begin{eqnarray}\label{omegaBall}
	\omega_i^{\rm ballistic} &\approx& \omega_i -\frac{M_2\omega_2}{M_i\omega_i}\frac{\gamma}{2} - i\frac{M_2}{M_i}\frac{\Gamma}{2}
\end{eqnarray}
where we evaluated both the frequency shift (real part) and the decay rate (imaginary part) to lowest order in the interaction strength $a$.
Both the frequency shift and the decay rate are much smaller than $\delta\omega$ in the regime where Eq.~\eqref{omegaBall} is valid.
For low temperatures, $T \ll \epsilon_{F,1}$, one can use Eq.~\eqref{lowTlimits} to obtain for the frequency shift of the order of
\begin{equation}\label{omegashiftSmalla}
	\Delta \omega_i \sim \frac{N_2}{N_i}\, k_{F,1} a\, \bar\omega \ll \delta \omega
\end{equation}
while the decay rate of the oscillations is essentially given by the single-particle scattering rate,
\begin{eqnarray}
	\frac{1}{\tau_{\rm osc,i} }
	&\sim& \frac{N_2}{N_i} \frac{(k_B T)^2}{\hbar^2\, \epsilon_{F,1}}\, (k_F a)^2  \ll \delta \omega.
\end{eqnarray}
For high temperatures, $T\gg\epsilon_{F,1}$, the frequency shift drops faster than the decay rate and is therefore difficult to observe.

In appendix~\ref{appendix-sigma-singular} we show that in the ballistic regime the memory matrix formalism does not reproduce a dephasing of oscillations which gives rise to an extra effective decay rate linear in the scattering length 
$a$. This failure of the approach can be traced back to the fact that in the limit $a \to 0$ an infinite set of further slow modes exists which we did not include into the set of slow modes $A_n$, see appendix~\ref{appendix-sigma-singular} for details.

\begin{figure}
	\includegraphics[width=\columnwidth]{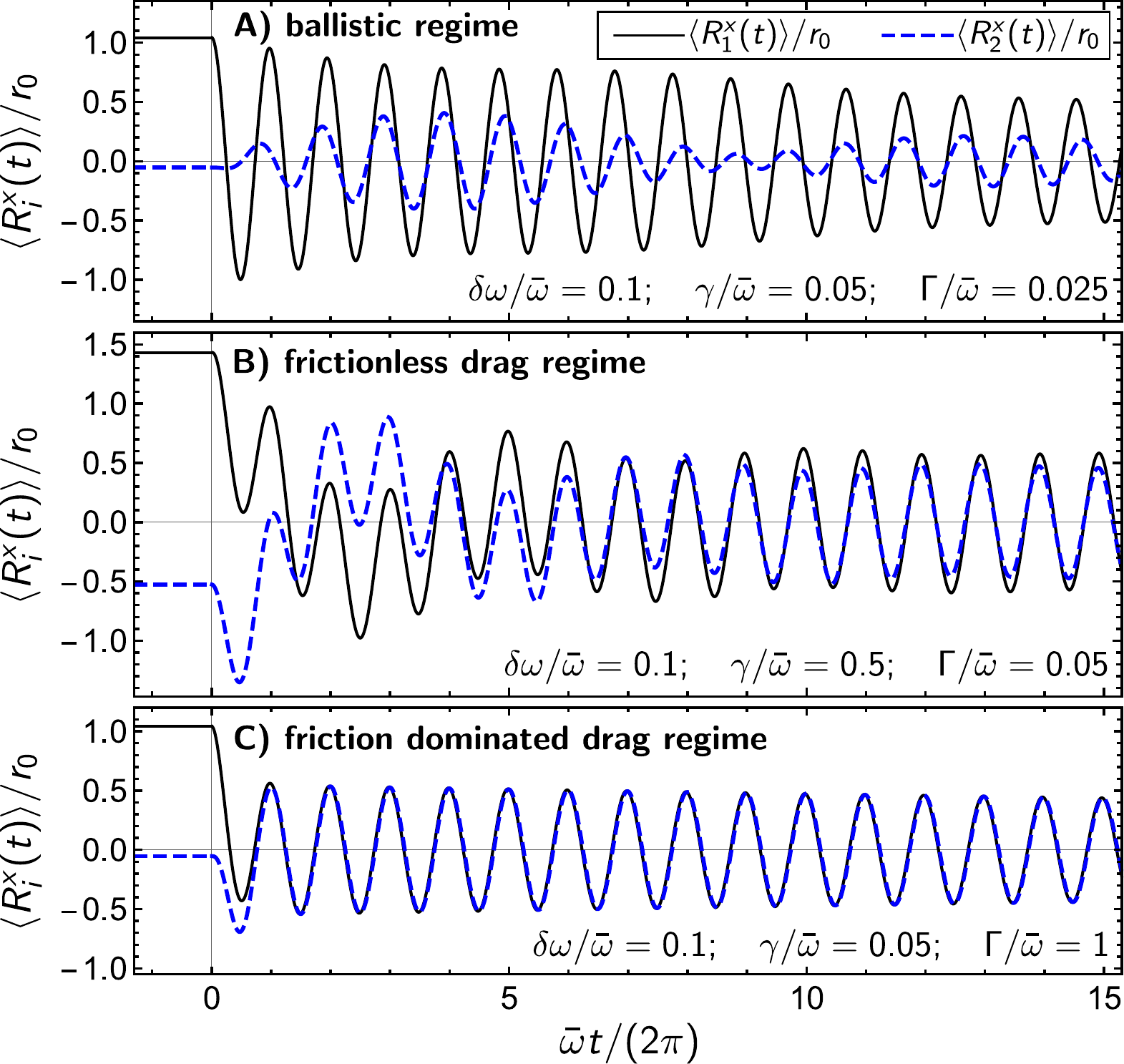}
\caption{\label{figRealTime}
Response to a constant displacement $\vec r_1^0(t<0)=r_0\hat{\vec e}_x$ of the trap potential for species $1$ that is switched of suddenly at time $t=0$.
The solid black (dashed blue) line shows the expectation value $\langle R_1^x(t)\rangle$ ($\langle R_2^x(t)\rangle$) of the center position of the first (second) atomic cloud, respectively, see Eq.~\eqref{realTime}.
The calculations were done for $N_1=N_2$, $m_1=m_2$, $\frac{\delta\omega}{\bar\omega}=0.1$ and $\gamma$ and $\Gamma$ as specified for each case.
For high friction $\Gamma$ (last plot), the oscillations of the two species synchronize quickly despite the finite difference $\delta\omega$ of the respective trapping frequencies, and the remaining COM oscillation decays only slowly on the time scale $\Gamma/\delta\omega^2$, see Eq.~\eqref{omegaFdd}.
}
\end{figure}
In a cold-atom experiment, one can directly observe the response of the clouds in real time.
From the theory side, the real-time response can be obtained by Fourier transformation of the susceptibility, Eq.~(\ref{chiMemoryMatrix}).
We consider the following setup: for time $t<0$ a constant force is applied to the first species.
Equivalently, we set in the Hamiltionian (Eq.~\eqref{hamiltonian}), $\vec r^0_2(t)=0$ and $\vec r^0_1(t)=\vec r_0= r_0 \hat{\vec e}_x$ with $r_0>0$ for $t<0$ ($\hat{\vec e}_x$ is the unit vector in $x$ direction).
The force is suddenly switched off, $\vec r^0_i(t)=0$, for $t\ge 0$.
In Fig.~\ref{figRealTime} the expectation value 
\begin{eqnarray}
\langle \vec R_i(t)\rangle=  M_1 \omega_1^2\vec r_0 \int_{-\infty}^0 \chi_{i1}(t-t') dt' \label{realTime}
\end{eqnarray}
is plotted as a function of time for both species, $i=1,2$.
In Fig.~\ref{figRealTime}A an example from the ballistic regime is shown.
Due to the finite interactions the two modes couple and a beating pattern emerges which is characteristic for the superposition of the two frequencies $\omega_1$ and $\omega_2$.
All oscillations decay on a time scale set by $1/\Gamma$.

\subsection{Frictionless drag regime}

In the ballistic regime, the approximate symmetry which protects COM oscillations is of no relevance. This is different
in cases where interactions are sufficiently strong so that the first species drags the second one either directly
by the Hartree potential (frictionless drag) or by dissipative processes (friction dominated drag).
In these regimes, the eigenmodes are characterized by a long-lived mode of COM oscillations, where both atomic clouds oscillate in parallel, and a mode of relative oscillations, which decays more quickly.
If all particles synchronize their oscillation, then one can expect that  the COM mode is approximately described by undamped oscillations of a rigid body of total mass $M_{\rm tot}=M_1+M_2$ oscillating in an effective potential  $\frac{1}{2} (M_1 \omega_1^2+M_2 \omega_2^2)r^2$.
The oscillation frequency in this limit is given by
\begin{equation}\label{omegaCom0}
	\omega_{\rm COM}^{(0)} = \sqrt{\frac{M_1 \omega_1^2+M_2 \omega_2^2}{M_{\rm tot}}}.
\end{equation}

We first consider the frictionless drag regime, which is reached when the  frequency shift described by Eq.~\eqref{omegaBall} becomes larger than the difference $\delta\omega$ of the trapping frequencies, $|\gamma|\gg \delta\omega$, and at the same time interaction effects between the two species are dominated by the effective potential rather than scattering, i.e., $|\gamma| \gg \Gamma$.
For low $T$, this regime is obtained for $\frac{\delta \omega}{\omega} \ll k_F a \ll \frac{\hbar \omega \, \epsilon_{F1}}{(k_B T)^2}$.

In this frictionless drag regime we can use perturbation theory in $\delta \omega$ to calculate the frequency shift and lifetime of the COM oscillations. We obtain 
\begin{equation}\label{omegaFld}
	\omega_{\rm COM} \approx \omega_{\rm COM}^{(0)} + \frac{2M_1^2M_2}{M_{\rm tot}^3} \left(\frac{\delta\omega^2}{\gamma} - i \frac{\delta\omega^2\Gamma}{\gamma^2}\right).
\end{equation}

As expected, $\omega_{\rm COM} \to \omega_{\rm COM}^{(0)}$ for large $\gamma\sim a$, as the increasing drag effect causes the two atomic clouds to oscillate more and more in parallel despite the small difference $\delta\omega$ of their trapping frequencies.
Defining $\Delta \omega$ by the shift relative to  $\omega_{\rm COM}^{(0)}$, we obtain for low $T$
\begin{equation}
	\Delta \omega \sim  \frac{ N_2}{N_{1}} \frac{1}{ k_F a}\frac{\delta\omega^2}{\omega} \ll \delta \omega
\end{equation}
where we used again Eq.~\eqref{lowTlimits}.
While the frequency shift is proportional to $1/k_F a$, the lifetime turns out to be independent of the interaction strength in this regime,
\begin{equation}
	\frac{1}{\tau_{\rm COM}} \sim  \frac{ N_2}{N_{1}} \left( \frac{\delta\omega}{\omega}\right)^2 \frac{(k_B T)^2}{\hbar^2\, \epsilon_{F,1}} \ll \Delta \omega \ll \delta \omega
\end{equation}
Note that both $\Delta \omega$ and $\frac{1}{\tau_{\rm COM}} $ are proportional $(\delta \omega)^2$ as
frequency shift and decay only arise from the small contributions violating the symmetry which approximately protecs COM oscillations.

For completeness, we mention that the complex frequency of the mode of relative oscillations \cite{bruun} is given by
\begin{equation}
	\omega_{\rm rel} \approx \omega_{\rm COM}^{(0)} - \frac{M_{\rm tot}}{2M_1} (\gamma + i\Gamma). \label{decayRelative}
\end{equation}
This mode is damped by the single-particle relaxation time $\Gamma$ and obtains a large frequency shift of the order of $k_F a \, \omega$ for low $T$. As discussed above, the formula above ignores extra dephasing effects, see Appendix \ref{appendix-dephasing}.

In Fig.~\ref{figRealTime}B the real-time response is shown in the frictionless drag regime using again  Eq.~(\ref{realTime}).
Due to the strong repulsive interactions the two clouds repel each other such that $\langle R_1^x(t\leq0)\rangle$ is larger than $r_0$ and $\langle R_2^x(t\leq0)\rangle$ is negative.
After the external force has been switched off at $t=0$, the first cloud moves towards the center, first pushing the second cloud further away.
After some time, the relative motion of the two clouds has decayed, the oscillations lock into each other and only the COM oscillations remain.
The decay of the latter is given by the tiny rate $\sim \Gamma ({\delta \omega}/{\gamma})^2$, see Eq.~(\ref{omegaFld}), due to the approximate symmetry.

\subsection{Friction dominated drag regime}

Experimentally, the most important regime is perhaps the hydrodynamic regime, where friction dominates, $\Gamma \gg |\gamma|, \delta \omega$.
For $k_B T \ll \epsilon_{F1}$, this condition is fulfilled for $k_F a \gg \frac{\hbar \omega \, \epsilon_{F,1}}{(k_B T)^2}$ and $k_F a \gg \sqrt{\frac{\hbar\delta \omega \, \epsilon_{F1}}{(k_B T)^2}}$, which is, e.g., realized with realistic experimental parameters of $k_{F,1}a\approx 0.2$, $N_1 \approx N_2 \approx 10^6$, $\frac{\delta\omega}{\bar\omega} \approx 0.1$, and $k_B T \approx 0.3\epsilon_{F,1}$ (the Fermi energies are given by $\epsilon_{F,i}=\hbar\omega_i(6N_i)^{1/3}$).
Note that this regime is always reached in the thermodynamic limit defined by $N_i \to \infty$, $\omega_i \to 0$ with $\epsilon_{F,i} = const$.
Furthermore, we demand as above that $ k_F a \ll 1$.

The complex eigenfrequency of the COM mode is again obtained from perturbation theory in $\delta \omega$ and has the form 
\begin{equation}\label{omegaFdd}
	\omega_{\rm COM} \approx \omega_{\rm COM}^{(0)} + \frac{2M_2^3}{M_{\rm tot}^3} \left(\frac{\delta\omega^2\gamma}{\Gamma^2} - i \frac{\delta\omega^2}{\Gamma}\right).
\end{equation}
Similar to the frictionless drag regime, interaction effects are suppressed for large $\Gamma$ as the friction synchronizes the oscillations of the two atomic clouds. For low $T$ we obtain the decay rate
\begin{equation}
	\frac{1}{\tau_{\rm COM}} \sim \frac{1}{(k_{F,1} a)^2} \left( \frac{\hbar \delta\omega}{k_B T}\right)^2 \frac{\epsilon_{F,1}}{\hbar} \ll \delta \omega
\end{equation}
The frequency shift  is in this regime much smaller than the decay rate,
\begin{equation}
	\Delta \omega \ll \frac{1}{\tau_{\rm COM}}
\end{equation}
and therefore difficult to observe.
For low $T$ one obtains $\Delta \omega \sim \frac{\hbar^2 (\delta \omega)^2 \epsilon_{F1}^2 \bar\omega}{(k_B T)^4 (k_{F,1} a)^3}$.

Fig.~\ref{figRealTime}C demonstrates how efficient a large friction is to lock the motion of the two clouds into each other on a time scale set by $1/\Gamma$.
After this microscopic time-scale, only the center of mass oscillations remain, which decay very slowly on the time scale $\Gamma/(\delta \omega)^2$ , see Eq.~(\ref{omegaFdd}).
The motion of the two clouds is locked perfectly into each other.

\section{Numerical results\label{sectionNumericalResults}}

\subsection{Protocols and possible experimental setups}

\begin{figure}[t]
\centering
\includegraphics[width=\columnwidth]{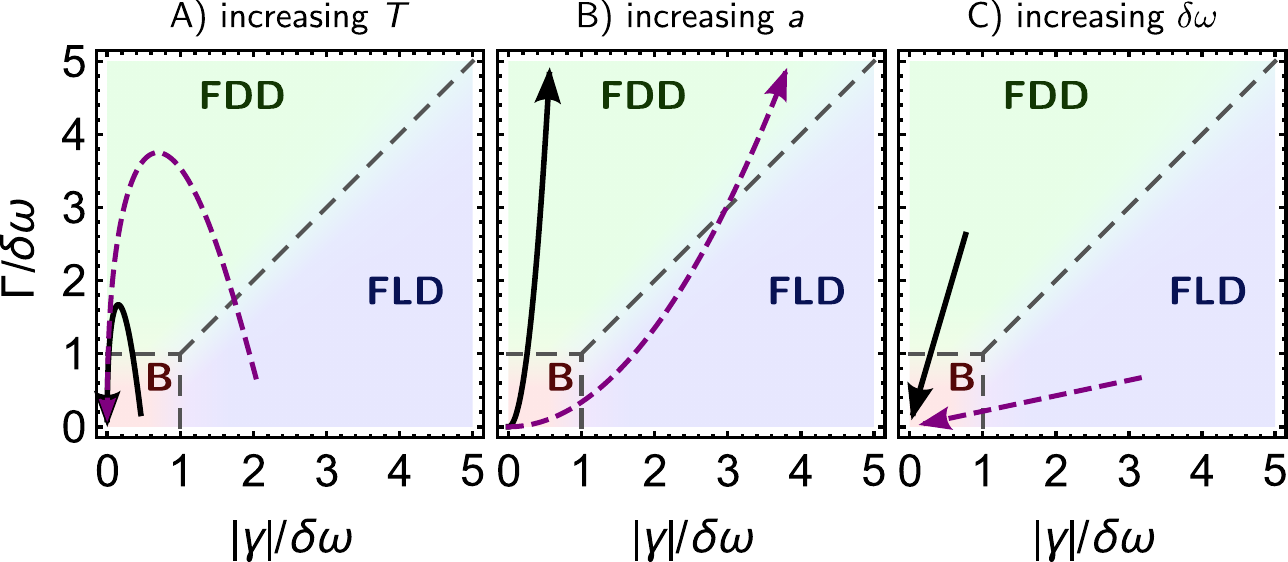}
\caption{\label{fig_parameterspace}
(Color online) Depending on which one of the quantities $\delta\omega$, $|\gamma|$, and $\Gamma$ is largest, the system is either in the ballistic regime (B), the frictionless drag regime (FLD) or the friction dominated drag regime (FDD).
The arrows show the trajectories of the system in the parameter space when the temperature, the interaction strength $a$, or the difference $\delta\omega$ of the trapping frequencies are increased.
All trajectories are calculated for $N_1=N_2=10^6$ and $m_1=m_2$.
A) $\frac{\delta\omega}{\bar\omega}=0.1$ ($0.01$), $k_{F,1}a=0.13$ ($0.06$), and $\frac{k_B T}{\epsilon_{F,1}}=0.05\ldots10$ ($0.07\ldots30$) for the solid black (dashed purple) trajectory, respectively.
B) $\frac{\delta\omega}{\bar\omega}=0.1$ ($0.01$), $\frac{k_B T}{\epsilon_{F,1}}=0.2$ ($0.1$), and $k_{F,1}a$ runs from $0$ to $0.025$ ($0.12$) for the solid black (dashed purple) trajectory, respectively.
C) $k_{F,1}a=0.1$ ($0.02$), $\frac{k_B T}{\epsilon_{F,1}}=0.2$ ($0.1$), and $\frac{\delta\omega}{\bar\omega}=0.03\ldots0.4$ ($0.002\ldots0.03$) for the solid black (dashed purple) trajectory, respectively.
}
\end{figure}

Depending on the setup of the  cold-atom experiment, there exist various possibilities to access the different physical regimes described in section \ref{sectionAnalyticResults}.
First, by changing the cooling protocol, it is possible to access a broad range of temperatures. Second, by using an Feshbach resonance one can tune the scattering length.
Third, if one has an experimental realization where the trapping potential of the two species can be varied independently, one can directly tune $\delta \omega$.
In Fig.~\ref{fig_parameterspace} we show how each of these methods leads to a different trajectory in the parameter space spanned by $\gamma/\delta \omega$ and $\Gamma/\delta \omega$.

\begin{figure}[tbp]
\centering
\includegraphics[width=\columnwidth]{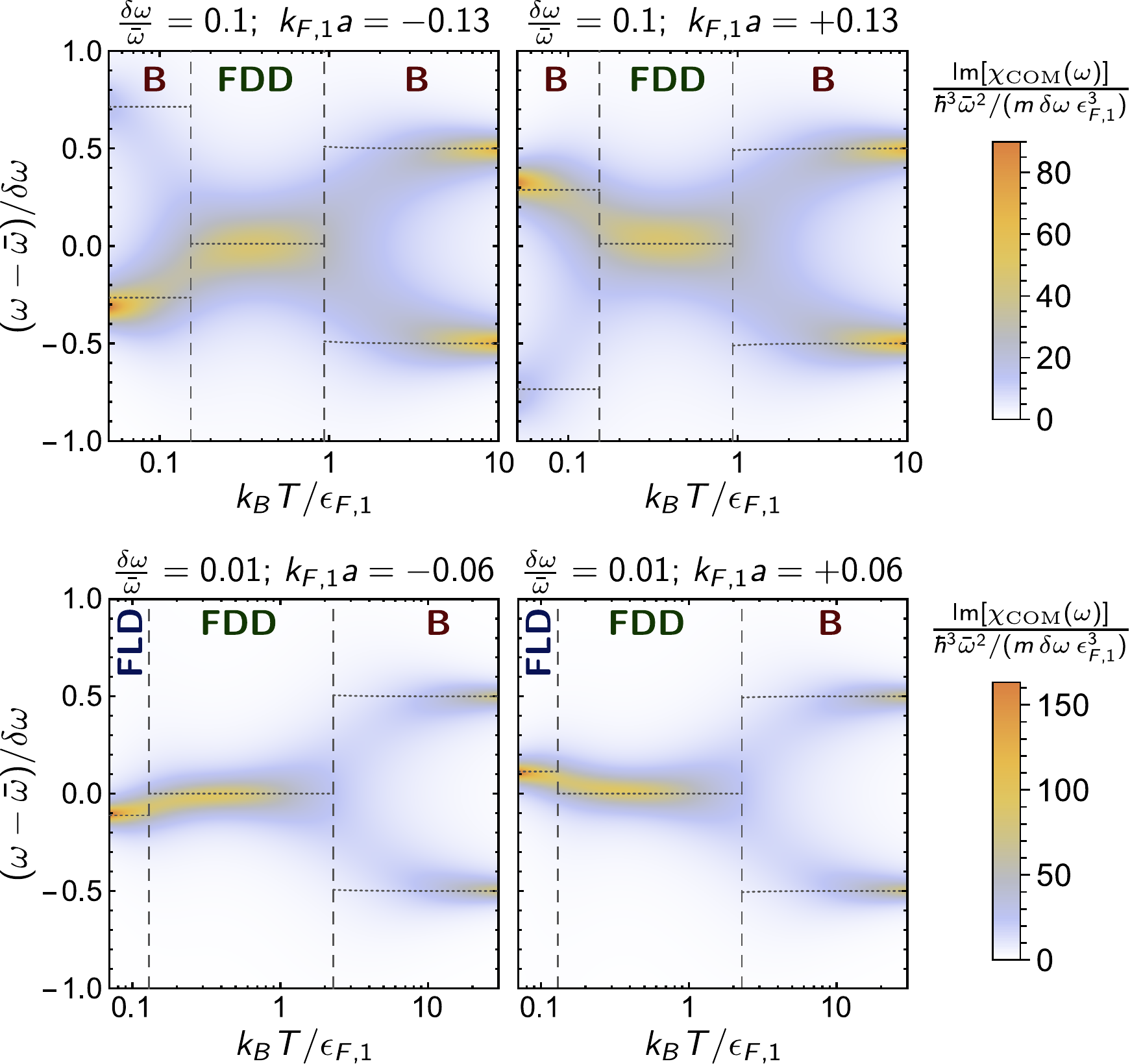}
\caption{\label{fig_tscan}
(Color online) Imaginary part of $\chi_{\text{COM}}(\omega)$, as defined in Eq.~\eqref{chicom}, as a function of temperature for attractive (left panels) and repulsive (right) interactions. The upper (lower) panels correspond to the solid  (dashed)  trajectories in Fig.~\ref{fig_parameterspace}A, respectively.
Dashed vertical lines indicate crossover temperatures where $\Gamma=|\gamma|$ or $\Gamma=\delta\omega$, horizontal dotted lines are the analytical predictions of Eqs.~\eqref{omegaBall}, \eqref{omegaCom0}, and \eqref{omegaFld} using Eqs.~\eqref{lowTlimits}--\eqref{highTlimits}.
For the top panels, the system evolves with increasing temperature from the ballistic (B) to the friction dominated drag (FDD) and back to the ballistic regime, while for the lower panels, the frictionless drag regime (FLD) is reached at low $T$.
Parameters: $N_1=N_2=10^6$, $m_1=m_2$, $\frac{\delta\omega}{\bar\omega}$ and $k_{F,1}a$ as stated above each plot.
}
\end{figure}

\begin{figure}[tbp]
\centering
\includegraphics[width=\columnwidth]{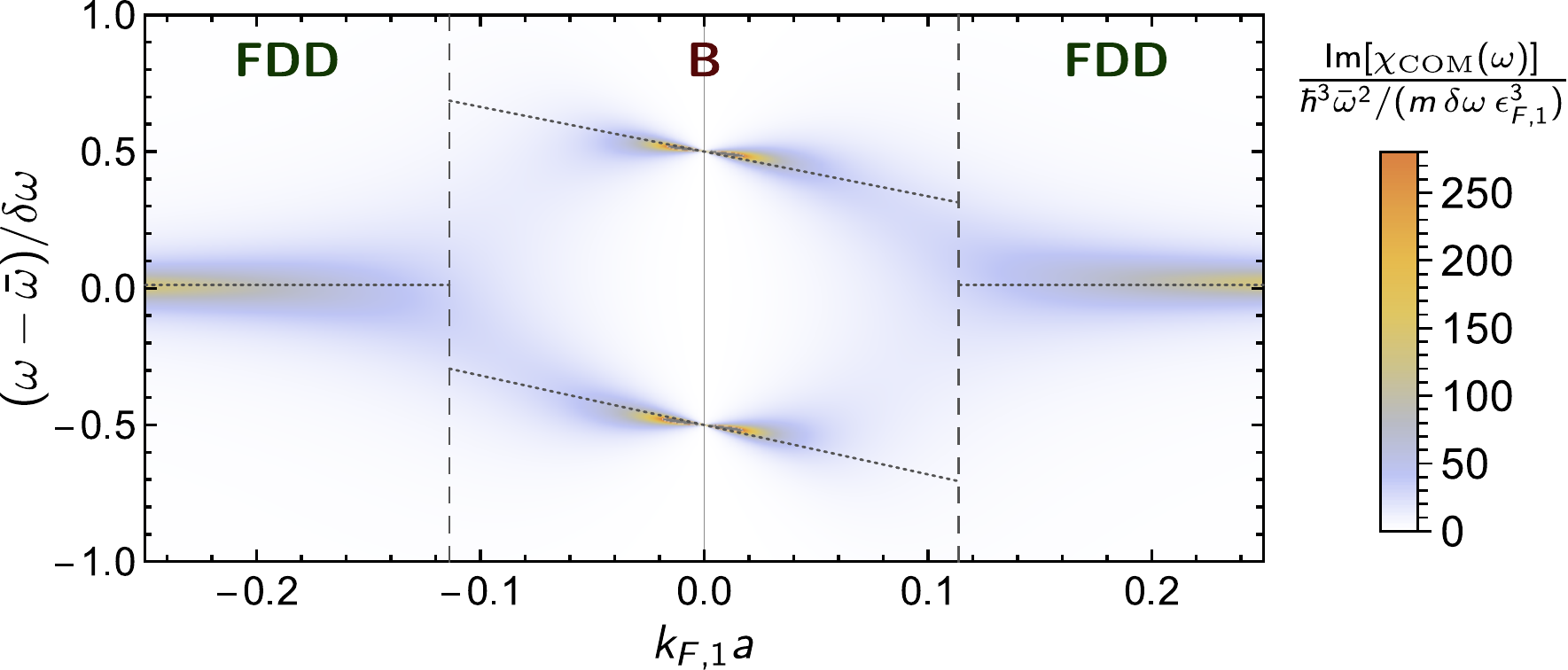}
\caption{\label{fig_ascan}
(Color online) $\text{Im}[\chi_{\text{COM}}(\omega)]$, Eq.~\eqref{chicom}, as a function of the scattering length $a$ (corresponding to solid line in Fig.~\ref{fig_parameterspace}B).
Dashed vertical lines indicate the scattering length where $\Gamma=\delta\omega$ and separate the ballistic (B) from the friction dominated drag regime (FDD).
The dotted lines are analytic predictions of the eigenfrequencies  based on Eqs.~\eqref{omegaBall} and \eqref{omegaFdd}, where we used the low-temperature limit, Eq.~\eqref{lowTlimits}, for the values of $\gamma$ and $\Gamma$.
Parameters: $N_1=N_2=10^6$, $m_1=m_2$, $k_B T=0.2\epsilon_{F,1}$,  $\frac{\delta\omega}{\bar\omega}=0.1$.
}
\end{figure}

To illustrate the various regimes, we will plot in the following sections, Figs. \ref{fig_tscan}, \ref{fig_ascan}, and \ref{fig_dwscan}, the imaginary part of 
\begin{eqnarray}\label{chicom}
	\chi_{\text{COM}}(\omega) = (1,1,0,0)\,\chi(\omega)\,(1,1,0,0)^T.
\end{eqnarray}
This describes the response of the center of mass to forces acting on both species simultaneously.
Experimentally, the susceptibility as function of frequency can, e.g., be obtained by observing the real-time dynamics followed by a Fourier transformation.

\begin{figure}[tbp]
\centering
\includegraphics[width=\columnwidth]{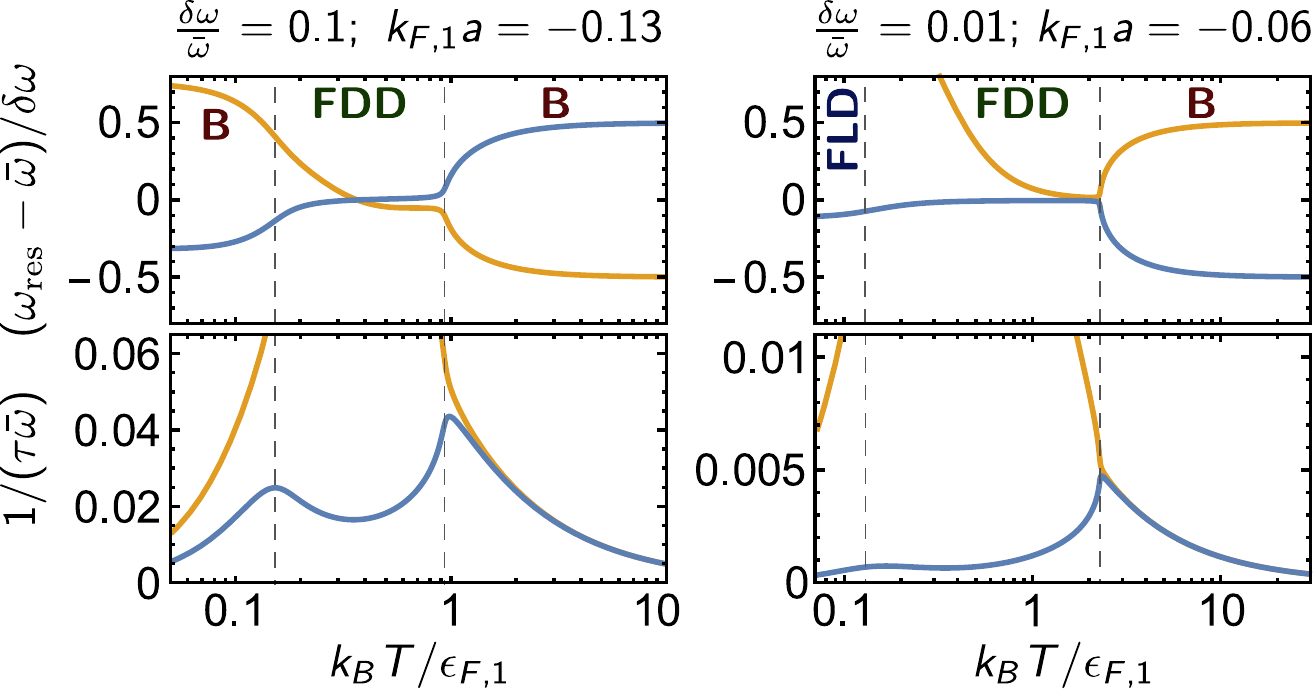}
\caption{\label{fig_decayrates}
(Color online) Resonance frequencies $\omega_{\text{res}}$ and decay rates $1/\tau$ of the two eigenmodes of the system as a function of temperature, calculated from the real and imaginary part of the eigenvalues of $\Omega-i\Sigma$, respectively, see Eq.~\eqref{chiMemoryMatrix}.
In all four panels, the blue graph corresponds to the mode with longer life-time $\tau$.
Dashed vertical lines indicate crossover temperatures between the ballistic (B), the frictionless drag (FLD), and the friction dominated drag (FDD) regime.
The left (right) column corresponds to the top (bottom) panel in the left column of Fig.~\ref{fig_tscan} and to the solid (dashed) trajectory in Fig.~\ref{fig_parameterspace}A, respectively.
Parameters: $N_1=N_2=10^6$, $m_1=m_2$.
}
\end{figure}
\subsection{Increasing the temperature}

While $\gamma$ decreases monotonically as a function of temperature, $\Gamma$ vanishes for both $T\to0$ and $T\to\infty$ and has a maximum at $k_B T \sim \epsilon_{F,1}$, see  Fig.~\ref{fig_gammaplot} and Eqs.~\eqref{lowTlimits}--\eqref{highTlimits}.
Therefore, two scenarios are possible when one increases $T$ while keeping all other parameters constant.
If interactions are weak, $k_{F,1}|a| \ll \frac{\delta\omega}{\bar\omega}$ (solid black trajectory in Fig.~\ref{fig_parameterspace}A), then the system is in the ballistic regime for low temperatures, may reach the friction dominated drag regime at intermediate temperatures $k_B T\sim \epsilon_{F,1}$ provided that $k_{F,1} |a|\gg \sqrt{\frac{\hbar\delta\omega}{\epsilon_{F,1}}}$ and returns to the ballistic regime for high temperatures.
If, on the other hand $k_{F,1}|a| \gg \frac{\delta\omega}{\bar\omega}$ (dashed purple trajectory in Fig.~\ref{fig_parameterspace}A), then the system is in the frictionless drag regime at low temperatures.
Increasing the temperature to the order of the Fermi energy will typically drive the system into the friction dominated drag regime unless $k_{F,1}|a| \ll \hbar\bar\omega/\epsilon_{F,1} \sim N_1^{-1/3}$.
At high temperatures, the ballistic regime is always realized.

For a quantitative analysis, we consider two concrete systems corresponding to the two trajectories in Fig.~\ref{fig_parameterspace}A.
In both cases, $N_1=N_2=10^6$ and $m_1=m_2$.
In the first system, $\frac{\delta\omega}{\bar\omega}=0.1$ and $k_{F,1}|a|=0.13$, while in the second case we use $\frac{\delta\omega}{\bar\omega}=0.01$ and $k_{F,1}|a|=0.06$.
We evaluate the integrals in Eqs.~\eqref{gammaexact} and \eqref{Gammaexact} numerically for these two systems.
Due to the spherical symmetry of the dispersion relation and the trapping potentials, the 12-dimensional integral in Eq.~\eqref{Gammaexact} can be reduced to a five-dimensional integral, which we evaluate using a Monte Carlo integration.

The different regimes can clearly be identified in plots of $\text{Im}[\chi_{\text{COM}}(\omega)]$, Eq.~(\ref{chicom}), shown in Fig.~\ref{fig_tscan}, describing excitations of the COM motion. The vertical dashed lines in  Fig.~\ref{fig_tscan}
correspond to the crossovers from one regime to the other, see Fig.~\ref{fig_parameterspace}A, while the horizontal dotted lines give the analytical predictions for oscillation frequencies.
The upper two (lower two) plots  in Fig.~\ref{fig_tscan} correspond to the solid (dashed) line in  Fig.~\ref{fig_parameterspace}A. On the left side, we consider attractive, on the right side repulsive interactions. 

The ballistic regime is characterized by the presence of two peaks: the two clouds oscillate independently with different frequencies. In contrast, a single peak located approximately at $\omega_{\rm COM}\approx \bar \omega$
characterizes the two drag regimes where the oscillation of the two clouds synchronizes. A second (much broader) mode describing relative oscillations does not show up
as for $\chi_{\rm COM}(\omega)$ we only consider a situation where both clouds are displaced
in the same direction (see Fig.~\ref{fig_decayrates} for a plot of both resonance frequencies as a function of temperature).
Note that for the chosen paramters, the system is not very deep in the ballistic regime for low $T$. This does not only lead to considerable shifts of the oscillation frequencies (see below) but also affects the weight of the two modes: the mode which is in frequency closer to $\omega_{\rm COM}^{(0)}$ clearly dominates.

In the low-temperature ballistic regime, the interactions increase (decrease) the oscillation frequencies as the curvature of the potential increases (decreases) due to the attractive (repulsive) interaction with the other species, respectively.
Interestingly, the effect is opposite for the drag-dominated regimes, best visible for the low-temperature regime in the lower two panels of Fig.~\ref{fig_tscan}.
This higher-order effect, well described by our analytical formulas (\eqref{omegaFld} and \eqref{omegaFdd}, drawn as dotted lines in Fig.~\ref{fig_tscan}),
arises from level repulsion from the mode of relative oscillations.

Fig.~\ref{fig_decayrates} shows the (real part of the) resonance frequencies and the decay rates $1/\tau$ of both eigenmodes of the system for the two cases corresponding to the left panels in Fig.~\ref{fig_tscan}. The maxima of the decay rate of the long-lived mode (lower curve in the lower panels of Fig.~\ref{fig_decayrates}) trace the crossover from one regime to the next. The minimum in the friction dominated drag regime, where $1/\tau_{\text{COM}}$ is proportional to the inverse of the single-particle scattering rate $\Gamma$, thereby arises from the maximum of $\Gamma$ displayed in Fig.~\ref{fig_gammaplot}.
While, in the ballistic regime, both modes have a long life time, in the drag regimes, only one long-lived mode remains and the decay rate of the mode of relative oscillations shoots up.

\subsection{Increasing the interaction strength}

As $\gamma\propto a$ and $\Gamma\propto a^2$, the system evolves on a parabola in the parameter space of Fig.~\ref{fig_parameterspace}B when the interaction strength is increased.
While for weak interactions, the ballistic regime and for strong interactions the friction dominated drag regime is always realized, the frictionless drag regime is only reached if 
$\frac{\delta\omega}{\bar\omega}(k_BT/\epsilon_{F,1})^2 N_1^{1/3} \lesssim 0.03$.

Fig.~\ref{fig_ascan} shows numerical results for $\text{Im}[\chi_{\text{COM}}(\omega)]$, c.f.~Eq.~\eqref{chicom}, for the solid black trajectory from Fig.~\ref{fig_parameterspace}B.
Dotted lines are again analytic results of the eigenfrequencies.
The analytic prediction overestimates the slopes  of the eigenfrequencies in the ballistic regime since it was made based on the $T\to0$ limit of $\gamma$ given in Eq.~\eqref{lowTlimits}, while the actual value of $\gamma$ at $k_BT=0.2\epsilon_{F,1}$ is by a factor of $0.62$ smaller.

\begin{figure}[tbp]
\centering
\includegraphics[width=\columnwidth]{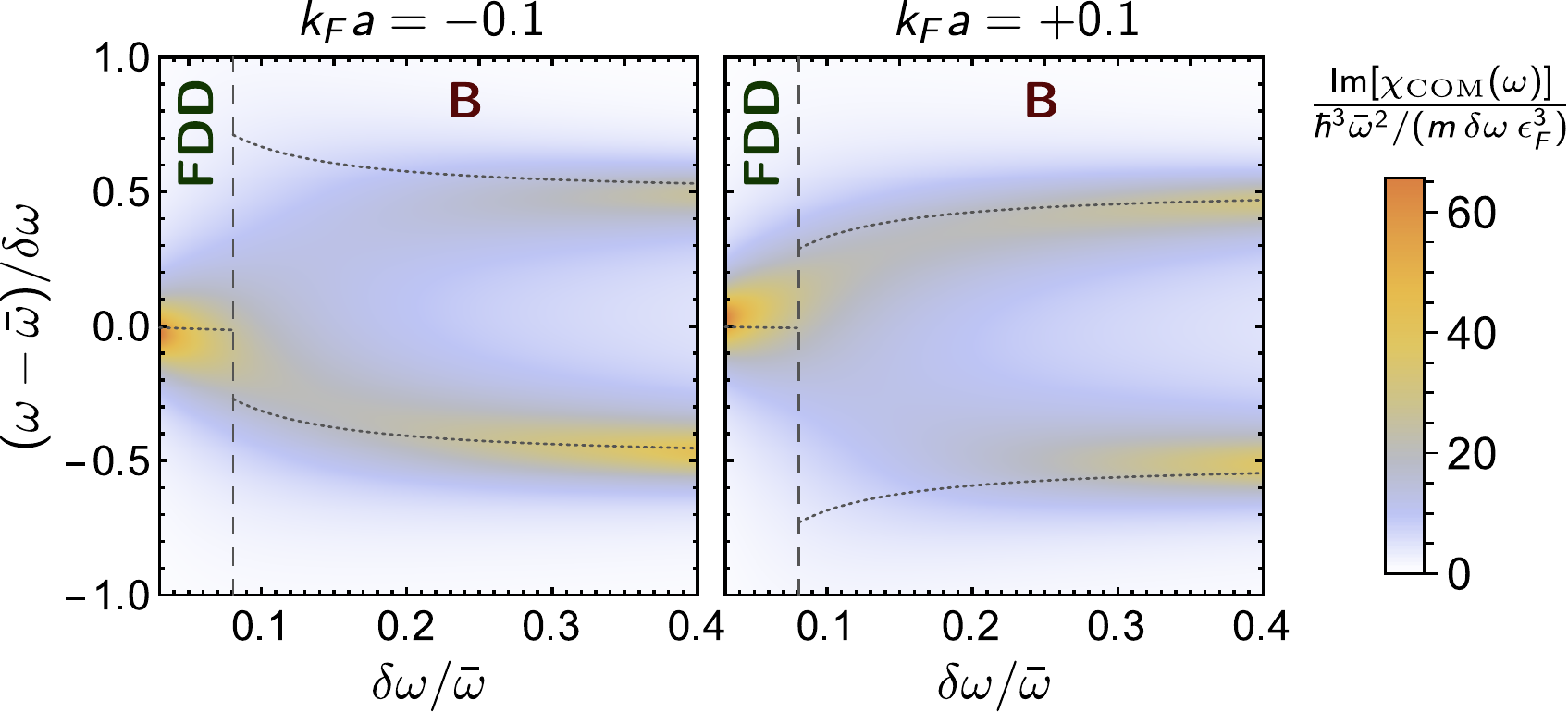}
\caption{\label{fig_dwscan}
(Color online) $\text{Im}[\chi_{\text{COM}}(\omega)]$, Eq.~\eqref{chicom}, as a function of the frequency difference $\delta\omega$ in the case $(k_BT/\epsilon_{F})^2 \,k_{F}|a| \,N_1^{1/3}=0.4>0.07$ (solid black trajectory in Fig.~\ref{fig_parameterspace}C).
At the dashed vertical line, $\delta\omega=\Gamma$, separating the friction dominated drag regime (FDD) from the ballistic regime (B).
The dotted lines are analytic predictions of the eigenfrequencies based on Eqs.~\eqref{omegaBall} and \eqref{omegaFdd}, where we used the low-temperature limit, Eq.~\eqref{lowTlimits}, for the values of $\gamma$ and $\Gamma$.
Parameters: $N_1=N_2=10^6$, $m_1=m_2$, $k_F a=\pm0.1$, $k_B T=0.2\epsilon_F$.
Here, $\epsilon_F$ and $k_F$ denote the Fermi energy and wave vector evaluated at $\delta\omega\to0$, respectively.
}
\end{figure}
\subsection{Increasing the frequency difference $\delta\omega$}

Since $\gamma$ and $\Gamma$ depend only weakly on $\delta\omega$ for $\frac{\delta\omega}{\bar\omega}\ll1$, the trajectories for increasing $\delta\omega$ in Fig.~\ref{fig_parameterspace}C are almost straight lines crossing at the origin.
For low temperatures and small $\delta\omega$, the friction dominated (frictionless) drag regime is realized if $(k_BT/\epsilon_{F,1})^2 \,k_{F,1}|a| \,N_1^{1/3}$ is larger (smaller) than $0.07$, respectively.
For large $\delta\omega$ (and weak interactions), the  system enters the ballistic regime.
Fig.~\ref{fig_dwscan} shows numerical results for $\text{Im}[\chi_{\text{COM}}(\omega)]$, c.f.~Eq.~\eqref{chicom}, corresponding to the solid black trajectory in Fig.~\ref{fig_parameterspace}C.
Dotted lines are analytic predictions of the eigenfrequencies based on the low-temperature limit, Eq.~\eqref{lowTlimits}.

\section{Conclusions}

The presence of approximate symmetries leads to a slow equilibration of a perturbed system.
We suggest that this physics can be studied with high precision experimentally by investigating
the center of mass oscillations of two species of ultracold atoms with different but similar mass.
Alternatively, one can also investigate, e.g., two spin species with the same mass but slightly different harmonic confinement. The mass difference and/or difference in the strength of the parabolic potential
breaks a dynamical symmetry which otherwise protects the center-of-mass oscillations from decay.

The interactions of the two species synchronizes the motion of the two clouds and thereby leads to a partial restoration of the dynamical symmetry: the interacting liquid can approximately be viewed as
having a single average mass and oscillating in a single average potential. As a consequence, the decay rate of the center-of-mass oscillations is strongly reduced and of the order of $\frac{(\delta \omega)^2}{\Gamma}$, where $\delta \omega$ is the difference of the trapping frequencies and $\Gamma$ the scattering-rate of the two species. Compared to other hydrodynamic modes (which can also have decay rates proprotional to the inverse of $\Gamma$) one obtains an extra reduction by the factor $(\delta \omega/\omega)^2$.

As all other modes have much faster decay rates, the approximate symmetry leads to an almost perfect drag of the two clouds: the center-of-masses  for each of the two species follow each other 
after a few scattering times.

For future investigations two directions are especially interesting: First, one can study the highly non-linear regime, where, for example, initially one species is separated far from the second one and one can study the evolution of the center-of-mass oscillations after the two clouds have violently crashed into each other in a setup similar to the one studied by the Zwierlein group \cite{sommer_universal_2011}. Second, one can investigate the interplay of  superfluidity and the approximate symmetry, which is of direct relevance for the experiments of the Salomon group \cite{ferrierbarbut_mixture_2014}. Here, in the center of the cloud and for small relative velocities, the superfluid components move without friction and only the normal components can scatter from each other.

\begin{acknowledgments}
We acknowledge useful discussions with E. Demler, J. Lux and C. Salomon and financial support from the DFG (SFB TR 12) and from Deutsche Telekom Stiftung.
\end{acknowledgments}

\bibliography{references}

\appendix

\section{Memory matrix formalism\label{appendix-memorymatrix}}
For a general set of observables $A_n$, time-dependent forces $f_n(t)$ on each observable are described by a contribution $H_{\text{ext}}=-\sum_n f_n(t)A_n$ to the Hamiltonian.
The response of some observable $A_m$ to the external forces $f_n(t)$ is described, to linear order in $f_n$, by the matrix of retarded susceptibilities $\chi_{mn}(\omega)$, as defined in Eq.~\eqref{def-chi},
via the relation
\begin{equation}
	\langle A_m(\omega)\rangle = 2\pi\delta(\omega)\langle A_m\rangle_{\text{eq.}} + \sum_n \chi_{mn}(\omega)f_n(\omega)
\end{equation}
where the Fourier transform of the external forces (and accordingly of the observable $A_m$) is defined by $f_n(\omega)=\int_{-\infty}^{\infty}e^{i\omega t}f(t)dt$ and $\langle\cdot\rangle$ denotes the expectation value in the perturbed system while $\langle\cdot\rangle_{\text{eq.}}$ is the equilibrium expectation value for $f_n(t)=0\;\forall t$.

To be specific, as described in the main text we study the dynamics of the ($x$-components of the) COM coordinates of two atomic clouds.
Thus, $A_n=(R_1^x,R_2^x,P_1^x,P_2^x)$ as defined in Eq.~\eqref{comcoordinates}.
The Hamiltonian, Eq.~\eqref{hamiltonian}, contains forces on the COM position coordinates $R_i^x$, $i=1,2$ of the two atomic clouds, which are given by $m_i\omega_i^2 r_i^{0,x}(t)$.
We do not consider forces on the COM momenta $P_i^x$, but nevertheless include $P_i^x$ in the set of operators $A_n$ since (i) the momenta can be observed in time-of-flight measurements and (ii) we expect excitations of the COM momenta to be long-lived in the regime $\frac{\delta\omega}{\bar\omega} \ll 1$ and the applied memory-matrix formalism requires a separation of time scales where the operators $A_n$ span the subspace of all slowly relaxing local observables.

We calculate the matrix of retarded susceptibilities $\chi_{mn}(\omega)$, Eq.~\eqref{def-chi}, by means of the memory-matrix formalism \citep{forster_hydrodynamic_1995,mori_transport_1965,zwanzig_ensemble_1960}.
In the following, we briefly review the central results of this technique.

A scalar product in the space of quantum-mechanical operators is defined by
\begin{equation}
	(A|B) := \int_0^\beta\!d\lambda\; \langle A^\dagger B(i\hbar\lambda)\rangle_{\text{eq.}} - \beta\langle A^\dagger\rangle_{\text{eq.}} \langle B\rangle_{\text{eq.}}
\end{equation}
where $\beta=1/(k_B T)$ is the inverse temperature and $B(i\hbar\lambda)=e^{-\lambda H}Be^{\lambda H}$ is the operator in the Heisenberg picture.
Instead of calculating the matrix of retarded susceptibilities $\chi_{mn}(\omega)$ directly, it is easier to first derive an expression for the matrix of retarded correlation functions $C_{mn}(\omega)$ defined by
\begin{equation}\label{cofomega}
	C_{mn}(\omega) = \int_{0}^{\infty}\!dt\; e^{i\omega t} \,(A_m(t)|A_n).
\end{equation}
It is easy to show that $\chi_{mn}(\omega)$ and $C_{mn}(\omega)$ are related via
\begin{equation}\label{relationcchi}
	\chi_{mn}(\omega) = i\omega\, C_{mn}(\omega) + (C_0)_{mn}
\end{equation}
where the entries of the equal-time correlation matrix $C_0$ are defined by
\begin{equation}\label{def-c0}
	(C_0)_{mn}=(A_m|A_n).
\end{equation}

Time evolution of an operator is described by $A(t)=e^{iLt}A$ with the Liouville (super\mbox{-})\nobreak\hspace{0pt}operator $L = \frac{1}{\hbar} [H,\,\cdot\,]$.
Thus, $ C_{mn}(\omega)$ is given by
\begin{alignat}{1}\label{cofomega2}
	C_{mn}(\omega) &= \int_{0}^{\infty}\!dt\; e^{i\omega t}(A_m|e^{-iLt}|A_n) \nonumber\\
	&= i\, (A_m|(\omega-L)^{-1}|A_n)
\end{alignat}
for $\text{Im}(\omega)>0$.

The operators $A_n$ span a subspace of the space of quantum-mechanical operators.
We define the projection (super-)operator $\mathcal P$ ($\mathcal Q$) onto (away from) this subspace by
\begin{equation}\label{def-projectors}
	\mathcal P = 1-\mathcal Q = \sum_{m,n} |A_m)\,(C_0^{-1})_{mn}\,(A_n|.
\end{equation}
Inserting $L=L\mathcal Q+L\mathcal P$ into Eq.~\eqref{cofomega2} and following some simple algebraic manipulations \cite{forster_hydrodynamic_1995} one arrives at a matrix equation for the retarded correlation functions,
\begin{equation}\label{cofomegamemory}
	C(\omega) = i\left(\omega - \Omega + i\Sigma(\omega)\right)^{-1} C_0
\end{equation}
where
\begin{alignat}{1}
	\Omega_{mn} &= i\sum_s (\dot A_m|A_s)(C_0^{-1})_{sn} \label{omegamatrix}\\
	\Sigma_{mn}(\omega) &= i \sum_s (\dot A_m| \mathcal Q (\omega-L\mathcal Q)^{-1}|\dot A_s)(C_0^{-1})_{sn} \label{memorymatrix}
\end{alignat}

The matrix $\Omega$ describes the evolution of the observables $A_n(t)$ if there was no coupling to any other degrees of freedom (i.e., if $L$ would commute with all $A_n$).
Effects due to the coupling of the $A_n$ modes to other modes are encoded in the memory matrix $\Sigma(\omega)$.

\section{Evaluation of the matrices $C_0$, $\Omega$, and $\Sigma(\omega)$\label{appendix-evaluation-sigma}}

In this section we evaluate the matrices $C_0$, $\Omega$ and $\Sigma(\omega)$, Eqs.~\eqref{def-c0},\eqref{omegamatrix}, and \eqref{memorymatrix}, for the model described by Eq.~\eqref{hamiltonian}.
All calculations are done perturbatively for small interaction strength $a$.
Scalar products are calculated in the local density approximation, which is valid for $N_1,N_2\gg1$.

\subsection{Equal-time correlation matrix $C_0$ \label{appendix-eval-c0}}

The equal-time correlation matrix $C_0$ is defined in Eq.~\eqref{def-c0}.
Due to their different signature under time reversal, the position and momentum operators have vanishing overlap, $(R^x_i|P^x_j)=0$.
We expand $C_0 \approx C_0^{(0)} + C_0^{(1)} + \mathcal{O}(a^2)$ for small $a$.
Without interactions, $a=0$, the two species decouple from each other, resulting in a diagonal matrix structure of $C^{(0)}_0$.
We obtain, within a local density approximation,
\begin{alignat}{1}\label{c00evaluated}
	C_0^{(0)} &= \begin{pmatrix}
		1/(M_1 \omega_1^2) & 0 & 0 & 0 \\
		0 & 1/(M_2 \omega_2^2) & 0 & 0 \\
		0 & 0 & M_1 & 0 \\
		0 & 0 & 0 & M_2
	\end{pmatrix}
\end{alignat}
where $M_i=N_im_i$.

In local density approximation, the momentum-momentum components of $C_0$ are not changed by interactions to first order in $a$.
Interactions only affect the position-position components and we obtain
\begin{equation}\label{c01evaluated}
	C_0^{(1)} = (R_1^x|R_2^x) \begin{pmatrix}
		-\frac{M_2\omega_2^2}{M_1\omega_1^2} & 1 & 0 & 0 \\
		1 & -\frac{M_1\omega_1^2}{M_2\omega_2^2} & 0 & 0 \\
		0 & 0 & 0 & 0 \\
		0 & 0 & 0 & 0
	\end{pmatrix}
\end{equation}
where
\begin{alignat}{1}
	(R_1^x|R_2^x) &\approx -\frac{1}{3\pi\hbar^4} \frac{a k_BT(m_1 m_2)^{3/2}}{N_1 N_2 m_{\text{red}}}
		\int_0^\infty\!\!\!dr\, r^4 g_1(r)g_2(r) \nonumber \\
	g_i(r) &= \text{Li}_{\frac12}\!\left(-e^{(\mu_i - \frac12 m_i\omega_i^2 r^2)/(k_BT)}\right). \label{r1r2scalarproduct}
\end{alignat}
Here, $\text{Li}_{\frac12}$ is the polylogarithm of order $\frac12$ and $\mu_i$ is the chemical potential for particles of species $i$ in the limit $a\to0$.
To arrive at the diagonal matrix elements of $C_0^{(1)}$ given in Eq.~\eqref{c01evaluated} one has to take into account that the actual chemical potentials depend on the interaction strength $a$.
Eq.~\eqref{c01evaluated} gives the result for fixed particle numbers, which was derived using the relation
\begin{equation}
	\left. \frac{\partial\, (R_i^x|R_i^x)}{\partial a} \right|_{N_i} = 
	\left. \frac{\partial\, (R_i^x|R_i^x)}{\partial a} \right|_{\mu_i}
	 - \frac{\partial\, (R_i^x|R_i^x)}{\partial N_i}
	\left. \frac{\partial N_i}{\partial a} \right|_{\mu_i}
\end{equation}
where the notation $|_x$ denotes that $x$ is kept constant in the derivative.
For the off-diagonal elements of $C_0$, corrections due to the dependency of the chemical potentials on the interaction strength are of higher order in $a$.

\subsection{Eigenfrequency matrix $\Omega$}

We expand $\Omega\approx\Omega^{(0)}+\Omega^{(1)}+\mathcal O(a^2)$ for small $a$.
The temporal derivatives $\dot A_m=\frac{i}{\hbar}[H,A_m]$ that appear on the right-hand side of Eq.~\eqref{omegamatrix} are given by
\begin{alignat}{1}
	\dot R^x_i &= P^x_i/M_i \label{heisenberg-eom}\\
	\dot P^x_i &= -M_i\omega_i^2 R^x_i + \frac{i}{\hbar}\left[H_{\text{int}}^{(12)},P^x_i\right]
	=: \dot P^x_{i,\text{trap}} + \dot P^x_{i,\text{int}} \nonumber
\end{alignat}
Setting $H_{\text{int}}^{(12)}=0$ in Eq.~\eqref{heisenberg-eom} and inserting into Eq.~\eqref{omegamatrix} using Eq.~\eqref{def-c0} one arrives directly at the leading-order contribution to the eigenfrequency matrix,
\begin{equation}\label{omega0}
	\Omega^{(0)} = \begin{pmatrix}
		0 & 0 & i/M_1 & 0 \\
		0 & 0 & 0 &i/M_2 \\
		-i M_1 \omega_1^2 & 0 & 0 & 0 \\
		0 & -i M_2 \omega_2^2 & 0 & 0
	\end{pmatrix}.
\end{equation}

Note that, once we set $\dot P^x_{i,\text{int}}=0$ in Eq.~\eqref{heisenberg-eom}, all scalar products that appear in the evaluation of $\Omega^{(0)}$ are canceled exactly by the factor $C_0^{-1}$ on the right-hand side of Eq.~\eqref{omegamatrix}.
Thus, all corrections to $\Omega^{(0)}$ due to interactions originate from the term $\dot P^x_{i,\text{int}}$ in Eq.~\eqref{heisenberg-eom}.
Since $(\dot P_{i,\text{int}}^x|P_j^x)=0$ due to different signature under time reversal, interactions only change the lower left $2\times2$ corner of the matrix $\Omega$.
We get, to leading order in $a$,
\begin{alignat}{1}
	(\dot P^x_{1,\text{int}}|R_1^x) &= -(\dot P^x_{2,\text{int}}|R_1^x) = -M_2\omega_2^2\,(R_1^x|R_2^x) \nonumber\\
	(\dot P^x_{2,\text{int}}|R_2^x) &= -(\dot P^x_{1,\text{int}}|R_2^x) = -M_1\omega_1^2\,(R_1^x|R_2^x). \label{pdotrscalarproduct}
\end{alignat}
Where $(R_1^x|R_2^x)$ is given in Eq.~\eqref{r1r2scalarproduct}.
Inserting Eqs.~\eqref{pdotrscalarproduct} into Eq.~\eqref{omegamatrix} yields the first order correction to the eigenfrequency matrix
\begin{alignat}{1}\label{omega1a}
	\Omega^{(1)} &= i(R_1^x|R_2^x) \begin{pmatrix}
		0 & 0 & 0 & 0 \\
		0 & 0 & 0 & 0 \\
		-M_2\omega_2^2 & M_1\omega_1^2 & 0 & 0 \\
		M_2\omega_2^2 & -M_1\omega_1^2 & 0 & 0
	\end{pmatrix} C_0^{-1}
\end{alignat}
Since the factor $(R_1^x|R_2^x)$ is already first order in $a$ we may approximate the matrix $C_0$ by $C_0^{(0)}$ (Eq.~\eqref{c00evaluated}).
This leads to
\begin{equation}\label{omega1}
	\Omega^{(1)} = -i M_1\omega_1^2 M_2\omega_2^2 \,(R_1^x|R_2^x) \begin{pmatrix}
		0 & 0 & 0 & 0 \\
		0 & 0 & 0 & 0 \\
		1 & -1 & 0 & 0 \\
		-1 & 1 & 0 & 0
	\end{pmatrix}
\end{equation}
Eqs.~\eqref{omega0}, \eqref{omega1}, and \eqref{r1r2scalarproduct} describe our result for the leading and next-to-leading order contribution to the eigenfrequency matrix $\Omega$.

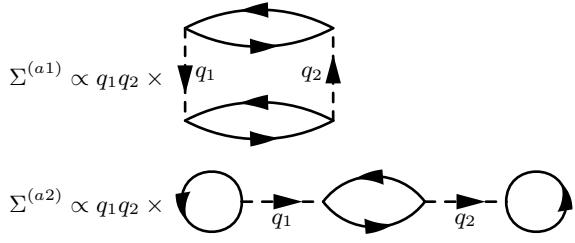
\begin{figure}
\begin{fmffile}{diagrams}
\begin{alignat*}{1}
	\Sigma^{(a1)} &\propto q_1 q_2 \times
	\begin{gathered}\hspace{-1mm}\begin{fmfgraph*}(70,35)
		\fmfleft{l1,l2}
		\fmfright{r1,r2}
		\fmf{fermion,right=0.25}{l1,r1,l1}
		\fmf{fermion,right=0.25}{l2,r2,l2}
		\fmf{dashes_arrow,label=$q_1$,l.dist=-12}{l2,l1}
		\fmf{dashes_arrow,label=$q_2$,l.dist=-12}{r1,r2}
	\end{fmfgraph*}\end{gathered} \\[3mm]
	\Sigma^{(a2)} &\propto q_1 q_2 \times
	\begin{gathered}\hspace{9mm}\begin{fmfgraph*}(100,30)
		\fmfleft{l}
		\fmfright{r}
		\fmf{dashes_arrow,label=$q_1$,l.dist=4.5}{l,v2}
		\fmf{fermion,right=0.5,tension=0.4}{v2,v3,v2}
		\fmf{dashes_arrow,label=$q_2$,l.dist=4.5}{v3,r}
		\fmffreeze
		\fmf{fermion,tension=0.9}{l,l}
		\fmf{fermion,tension=0.9}{r,r}
	\end{fmfgraph*}\hspace{9mm}\end{gathered}
\end{alignat*}
\end{fmffile}
\caption{\label{figDiagram}
The two types of diagrams contributing to $\Sigma^{(a)}$, Eq.~\eqref{sigmaa}.
Solid (dashed) lines represent fermions (inter-species interactions), respectively.
The second diagram would vanish in a homogeneous system due to momentum conservation at the vertices.
In a harmonic trap, however, $\Sigma^{(a2)}$ does not vanish and has poles at the trap frequencies (for vanishing intra-species interactions).}
\end{figure}
\subsection{Memory matrix $\Sigma(\omega)$\label{appendixSigma}}

The definition of the memory matrix $\Sigma(\omega)$ is given in Eq.~\eqref{memorymatrix}.
Note that one may insert an additional projection operator $\mathcal Q$ to the left of the vector $|\dot A_s)$ in the right-hand side of Eq.~\eqref{memorymatrix} without changing its value.
Using Eqs.~\eqref{heisenberg-eom} and the fact that $\mathcal Q$ projects onto the subspace of observables orthogonal to the space spanned by $|A_n)$ (see Eq.~\eqref{def-projectors}) we find $\mathcal Q|\dot R_i^x)=0=\mathcal Q|\dot P_{i,\text{trap}}^x)$.
Therefore, all contributions to $\Sigma(\omega)$ come from terms quadratic in $\dot P_{i,\text{int}}^x$ and thus at least of second order in the interaction strength $a$.
Neglecting higher order terms in $a$ we evaluate all scalar products on the right-hand side of Eq.~\eqref{memorymatrix} with respect to the non-interacting system and describe time evolution by the non-interacting Liouvillian $L_0=[H_0,\,\cdot\,]$.
Using further the fact that $L_0$ commutes with $\mathcal Q$ and that $C_0$ is diagonal (to lowest order in $a$) one finds that only the $P,P$-components of $\Sigma(\omega)$ have non-vanishing values given by
\begin{alignat}{1}
	\Sigma_{P^x_i,P^x_j}(\omega) &= i \frac{(\dot P^x_{i,\text{int}}| \mathcal Q (\omega-L_0)^{-1}|\dot P^x_{j,\text{int}})}{(P^x_j|P^x_j)} \nonumber\\
\end{alignat}
Inserting $\mathcal Q=1-\mathcal P$ leads to
\begin{alignat}{1}
	\Sigma_{P^x_i,P^x_j}(\omega)&= \Sigma^{(a)}_{P^x_i,P^x_j}(\omega) + \Sigma^{(b)}_{P^x_i,P^x_j}(\omega)
\end{alignat}
with
\begin{alignat}{1}
	\Sigma^{(a)}_{P^x_i,P^x_j}(\omega) &= i \frac{(\dot P^x_{i,\text{int}}| (\omega-L_0)^{-1}|\dot P^x_{j,\text{int}})}{(P^x_j|P^x_j)}\nonumber \\
&=\Sigma^{(a1)}_{P^x_i,P^x_j}(\omega)+\Sigma^{(a2)}_{P^x_i,P^x_j}(\omega) \label{sigmaa} \\
	\Sigma^{(b)}_{P^x_i,P^x_j}(\omega) &=- i \frac{(\dot P^x_{i,\text{int}}| \mathcal P  (\omega-L_0)^{-1}|\dot P^x_{\text{j,int}})}{(P^x_j|P^x_j)}.  \label{sigma-b}
\end{alignat}
Two types of diagrams, shown in  Fig.~\ref{figDiagram}, contribute to $\Sigma^{(a)}$, which we denote by  $\Sigma^{(a1)}$ and  $\Sigma^{(a2)}$ in the following. $\Sigma^{(a1)}$  describes how the scattering of quasi particles leads to momentum transfer from one species to the other. Evaluating the diagram within the local density approximation (i.e., by approximating the system locally by a homogeneous one) results 
in Eqs.~\eqref{mmatrix}--\eqref{Gammaexact} of the main text.

\subsection{Singular contributions to the memory matrix\label{appendix-sigma-singular}}

The discussion of the second diagram in Fig.~\ref{figDiagram},  $\Sigma^{(a2)}$, and of $\Sigma^{(b)}$ requires some more care.
While both contributions {\em vanish} within the local density approximation, they have both a {\em divergent} contribution
for $\omega=\omega_i$ if evaluated exactly (as long as intra-species interactions are absent, see below). We will argue that the divergencies cancel {\em exactly} if the full space of slow operator is considered. This is the reason why we ignore these contributions for our analysis despite the fact that the cancellation is only partial for the four modes considered by us.

First, we note that  $\Sigma^{(a2)}$ has a simple physical interpretation.
It describes that the momentum decays because, e.g., species 1 is affected by a single-particle Hartree potential
$V_H(\vec r)=\frac{4 \pi \hbar^2 a}{2 m_{\rm red}} \langle \Psi_2^\dagger(\vec r) \Psi_2(\vec r)\rangle$,  giving rise to an extra force  $\vec F_{H1}=-\int \frac{\partial V_H(\vec r)}{\partial \vec r}  \Psi_1^\dagger(\vec r) \Psi_1(\vec r) \,d^3 r$ contributing to
$\partial_t \vec P_1$. We obtain
\begin{alignat}{1}
\Sigma^{(a2)}_{P^x_i,P^x_j}(\omega)+\Sigma^{(b)}_{P^x_i,P^x_j}(\omega)  &= i \frac{(F_{Hi}^{x}| \mathcal Q (\omega-L_0)^{-1}|F_{Hj}^x)}{(P^x_j|P^x_j)}\label{divergent}
\end{alignat}
 In cases where we can approximate $V_H$ by a parabola, $\vec F_H$ is proportional to $\vec R_1$
and does not contribute to $\Sigma$ as $\mathcal Q  F^x_{Hi}=0$. 
Similar, if we include in our list of slow modes $A_n$ {\em all} operators which oscillate with frequency
$\omega_i$, then by construction $\Sigma^{(a2)}_{P^x_i,P^x_j}(\omega)+\Sigma^{(b)}_{P^x_i,P^x_j}(\omega)$ is non-singular for $\omega \to \pm \omega_i$. The fact that for our choice of slow modes an extra divergency remains, arises from a peculiar property of the harmonic oscillator. As all single-particle energy levels are equally spaced, there is an {\em infinite} number of hermitian operators oscillating with frequencies $\pm \omega_i$ in the non-interacting limit,
\begin{align}
A_{m,i,1}&=\int \Psi^\dagger_i(\vec r)r_x \hat h_i^m \, \Psi_i(\vec r)  \, d^3 r\\
A_{m,i,2}&=-i \int \Psi^\dagger_i(\vec r) \frac{\partial}{\partial r_x} \hat h_i^m \Psi_i(\vec r)  \, d^3 r
\end{align}
where $\hat h_i=-\frac{\hbar^2}{2 m_i} \frac{\partial^2}{\partial \vec r^2}+\frac{1}{2} m_i \omega_i^2 \vec r^2$ is the single particle Hamiltonian of species $i$.

In our analysis we have (i) only included the operators with $m=0$ and (ii) neglected the  divergent contributions discussed above. It is therefore important to ask 
to what extent our results are modified when further terms with $m>0$ are included. First, the accuracy of the result will increase as by construction the neglected terms become smaller and smaller.
It is important to note, that the more complicated operators $A_{m,i,j}$ with $m>0$ are {\em not} protected by any approximate symmetry.
Therefore their decay rate is {\em not} suppressed by factors of $(\delta \omega/\omega)^2$ and they will neither qualitatively nor quantitatively influence the final results in the limit where interactions become important (the hydrodynamic friction dominated drag regime).
Furthermore, in situations where intra-species interactions are present, these lead to a decay of  $A_{m,i,j}$ with $m>0$ but do not affect $A_{m,i,j}$ with $m=0$.
Therefore, if the decay rate due to intra-species interactions is sufficiently high, our results are again fully valid in all regimes considered.
Technically, this is reflected by the fact that the omitted terms (\ref{divergent}) are non-divergent if intraspecies interactions are included in $L_0$. 

In the ballistic limit where all friction can be ignored, however, our results presented in the main text 
miss a physically important effect: the decay of oscillations by dephasing (rather than decay by friction
considered by us). 
As we will show in the following section, this leads in the ballistic regime to a decay rate which is {\em linear} in the scattering length $a$ and is not covered in memory matrix approximations which neglect the modes
$A_{m,i,j}$ with $m>0$ and predict decay rates proportional to $a^2$.

\section{Decay by dephasing: a toy model\label{appendix-dephasing}}

In this section, we discuss the decay of COM oscillations for a simple toy model where non-interacting fermions scatter from a weak, smooth, and time-independent potential $V(r)$.
We will use the calculation to show that (i) the memory matrix approach correctly describes the average shift of frequencies due to the Hartree potentials but (ii) fails to reproduce the correct lifetime in the ballistic regime due to the problems discussed in section \ref{appendix-sigma-singular} above.

In the simplified model considered here, the atoms of the first species do not move while those of the second species oscillate in a potential given by the harmonic trap $\frac{1}{2}m_2\omega_2^2 r^2$ plus the static Hartree potential $V(r) = \frac{4\pi\hbar^2 a}{m_2} \langle \hat{n}_1(r)\rangle$, where $\langle \hat{n}_1(r)\rangle$ is the expectation value of the density of species $1$ in equilibrium. Formally, we consider the limit $m_1 \to \infty$, $m_1 \omega_1^2=m_2 \omega_2^2$,
$N_1=N_2 (m_1/m_2)^{3/2}$, $k_{F,2} a \ll  (m_2/m_1)^{3/2}$, such that the two clouds have the same shape $\langle \hat{n}_1(r)\rangle \approx  (m_1/m_2)^{3/2} \langle \hat{n}_2(r)\rangle$.

The imaginary part of the retarded susceptibility for the COM position of the second species is given by the Kubo formula,
\begin{alignat}{1} \label{imchiDephasing}
	\text{Im}[\chi_{R_2^x,R_2^x}(\omega)] =& \frac{\pi}{N_2^2} \sum_{\alpha,\alpha'} (f(\epsilon_{\alpha})-f(\epsilon_{\alpha'})) \, |\langle \alpha'|\hat{r}_x|\alpha\rangle|^2 \times \nonumber\\
	&\qquad\qquad \times \delta(\hbar\omega-(\epsilon_{\alpha'}-\epsilon_{\alpha})),
\end{alignat}
where $f$ is the fermi function and $|\alpha\rangle$ are single-particle eigenstates with energies $\epsilon_\alpha$.
We evaluate Eq.~\eqref{imchiDephasing} perturbatively for small $V$.
As the energy levels $\epsilon_n^{(0)}=\hbar\omega_2 (n+\frac32)$ of the unperturbed three-dimensional isotropic harmonic oscillator are degenerate, one has to diagonalize the matrix $\langle\alpha'|V|\alpha\rangle$ for each $n$-subspace.
Since $V(r)\propto\langle\hat{n}_1(r)\rangle$ is spherically symmetric, this is done by the states $|\alpha\rangle = |nlm\rangle$, where $l$ and $m$ are the quantum numbers of angular momentum.
To linear order in $V$, the eigenenergies of these states are independent of $m$ and given by $\epsilon_{n,l} = \epsilon^{(0)}_n + \langle nl|V|nl\rangle$, which we evaluate numerically by a  one-dimensional integration in the radial direction.
As we are only interested in the behavior of $\text{Im}[\chi_{R_2^x,R_2^x}(\omega)] $ for $\omega$ close to $\omega_2$ and to linear order in $V$, it is sufficient to calculate the matrix elements $\langle n'l'm'|\hat{r}_x|nlm\rangle$ to order $V^0$, which leads to the selection rules $n'=n\pm1$, $l'=l\pm1$, and $m'=m$  (with quantization axis in the $x$ direction).
We obtain
\begin{alignat}{1}
	&\sum_m |\langle n+1,l\pm1,m|\hat{r}_x|n,l,m\rangle|^2 = \nonumber\\
	&=\frac{\sqrt{\hbar}\,(2l+1) (2l+1\pm1) (2l+1\pm2) (2n+5\pm 2l\pm1)}{\sqrt{m_2\omega_2}\, 24 (2l\pm1) (2l+2\pm1)}.
\end{alignat}

\begin{figure}
\includegraphics[width=\columnwidth]{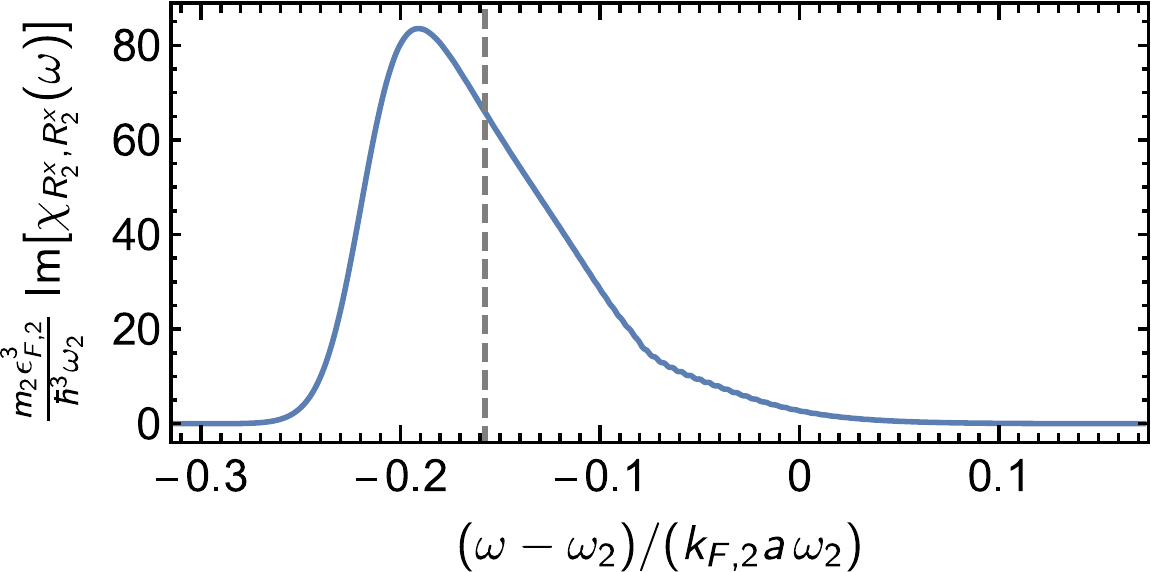}
\caption{\label{fig_dephasing}
Imaginary part of the retarded susceptibility for a non-interacting gas in a harmonic trap disturbed by a small extra potential $V(r) = \frac{4\pi\hbar^2 a}{m_2} \langle \hat{n}_1(r)\rangle$, see Eq.~\eqref{imchiDephasing}, where the $\delta$-function is approximated by a Gaussian with tiny standard deviation $\sigma_\omega=0.01|\Delta\omega|$.
The dashed line is the average shift $\Delta\omega$ of the peak position, Eq.~\eqref{domegaDephasingDef}, correctly predicted by the memory matrix method, Eq.~\eqref{domegaDephasingMM}.
Parameters: $k_BT/\epsilon_{F,2}=0.1$, $N_2=10^5$ in the limit $m_1/m_2 \to \infty$, $m_1 \omega_1^2=m_2 \omega_2^2$,
$N_1=N_2 (m_1/m_2)^{3/2}$, $k_{F,2} a \ll  (m_2/m_1)^{3/2}$.
}
\end{figure}

In Fig.~\ref{fig_dephasing} we show the resulting $\text{Im}[\chi_{R_2^x,R_2^x}(\omega)]$ for $\omega$ close to $\omega_2$.
The $\delta$-peak one would obtain for $V=0$ at $\omega=\omega_2$ is shifted linearly in the scattering length $a$ and also broadened {\em linearly} in $a$.
Note that the peak shape is {\em not} Lorentzian. The broading is not caused by inelastic scattering but arises instead from a simple dephasing effect: the frequency shifts linear in $V$ affect the energies of different eigenstates in a different way.
The dephasing linear in $V$ is {\em not} covered by the version of the memory matrix approach used by us, which does not take into account higher modes $A_{m,i,j}$ with $m>0$, and ignores divergent terms in $\Sigma(\omega)$ arising in a treatment beyond the local density approximation.
This is the main result of this section.
Note that this dephasing only affects the decay rates in the ballistic regime, Eq. (\ref{omegaBall}) and the decay rate of  the relative motion of the two species in the drag regimes, Eq.~(\ref{decayRelative}).
It is irrelevant in the friction dominated drag regime, where the inelastic scattering rate $\Gamma$ is much larger than the dephasing rate.
Furthermore, the dephasing effects are expected to be strongly reduced by the factor $(\delta \omega/\gamma)^2$ for the COM mode in the frictionless drag regime, cf.~Eq.~(\ref{omegaFld}), as for the synchronized motion of the two species the Hartree potential cancels to leading order.

In the following, we will show that our memory matrix approach does, however, correctly predict the {\em average} frequency shift, see dashed line in Fig.~\ref{fig_dephasing}.
The average frequency shift is defined by
\begin{equation} \label{domegaDephasingDef}
	\Delta \omega =\frac{1}{C} \int_{\omega_2-\delta}^{\omega_2+\delta} \frac{d\omega}{2\pi}\; (\omega-\omega_2) \, \text{Im}[\chi_{R_2^x,R_2^x}(\omega)]
\end{equation}
with the normalization $C=\int_{\omega_2-\delta}^{\omega_2+\delta} \frac{d\omega}{2\pi} \, \text{Im}[\chi_{R_2^x,R_2^x}(\omega)] = 1/(4N_2 m_2\omega_2)$ and $\delta<\omega_2$ chosen such that only the weight of the peak close to $\omega_2$ is captured.
Inserting Eq.~\eqref{imchiDephasing}, we find to linear order in $V$,
\begin{alignat}{1}
	\Delta \omega  &\approx  \frac{1}{\hbar N_2} \sum_{\alpha,\alpha'}(f(\epsilon^{(0)}_\alpha)-f(\epsilon^{(0)}_{\alpha'}))\; |\langle\alpha'|\hat{a}_x^\dagger|\alpha\rangle|^2 \times\nonumber\\
	&\qquad\qquad\qquad\times(\langle\alpha'|V|\alpha'\rangle - \langle\alpha|V|\alpha\rangle) \label{domegaDephasingDirect}
\end{alignat}
where $\hat{a}_x^\dagger$ is the ladder operator of the harmonic oscillator in $x$ direction.

Applying our version of the memory matrix, we obtain for the frequency shift to linear order in $V$ 
\begin{alignat}{1} \label{domegaDephasingMM}
	\Delta\omega &\approx - \frac{\omega_2}{2\hbar} (F_{2}^x|R_2^x) \nonumber\\
	&= \frac{1}{\hbar N_2} \sum_{\alpha,\alpha'} (f(\epsilon^{(0)}_\alpha)-f(\epsilon^{(0)}_{\alpha'}))
	 \langle \alpha'|\hat{a}_x^\dagger|\alpha\rangle \langle\alpha|[\hat{a}_x, V]|\alpha'\rangle
\end{alignat}
where $F_{2}^x = -\int \frac{\partial V}{\partial x} \Psi_2^\dagger(x)\Psi_2(x) \,d^3r$ is the force arising from the Hartree potential, and we used $\frac{\partial V}{\partial x} = \sqrt{2m_2\omega_2/\hbar}\,[\hat{a}_x,V]$ in the last equality.
We factorize the last matrix element in Eq.~\eqref{domegaDephasingMM} by inserting $1=\sum_{\tilde\alpha}|\tilde\alpha\rangle\langle\tilde\alpha|$ between the operators $\hat{a}_x$ and $V$, and write it explicitly in the eigenstates $|nlm\rangle$ of the harmonic oscillator,
\begin{alignat}{1}
	\langle\alpha|[\hat{a}_x, V]|\alpha'\rangle &= \sum_{\tilde n,\tilde l,\tilde m} \big(\langle nlm|\hat{a}_x|\tilde n\tilde l\tilde m\rangle \langle\tilde n\tilde l\tilde m|V|n'l'm'\rangle \nonumber\\
	&\qquad - \langle nlm|V|\tilde n\tilde l\tilde m\rangle \langle\tilde n\tilde l\tilde m|\hat{a}_x|n'l'm'\rangle \big). \label{dephasingMatrixElement}
\end{alignat}
Inserting Eq.~\eqref{dephasingMatrixElement} into Eq.~\eqref{domegaDephasingMM}, using the properties $\langle nlm|V|\tilde n\tilde l\tilde m\rangle \propto \delta_{l,\tilde l}\,\delta_{m,\tilde m}$ for the spherically symmetric potential $V$ and $\langle nlm|\hat{a}_x|\tilde n\tilde l\tilde m\rangle \propto \delta_{n+1,\tilde n}$, reproduces exactly the average frequency shift, Eq.~\eqref{domegaDephasingDirect}, derived from the direct calculation of $\text{Im}[\chi_{R_2^x,R_2^x}(\omega)]$.

\end{document}